\documentclass[english, preprint]{aastex63}
\usepackage[T1]{fontenc}
\usepackage[latin9]{inputenc}
\usepackage{bm}
\usepackage{amstext}
\usepackage{graphicx}
\usepackage{amsmath}
\usepackage{algorithm}
\usepackage[noend]{algpseudocode}

\makeatletter
\providecommand{\tabularnewline}{\\}
\makeatother

\usepackage{babel}
\begin{document}
\global\long\def\grad{\bm{\nabla}}%
\global\long\def\vgsh#1#2{\mathbf{Y}_{#2}^{#1}}%
\global\long\def\vgshlm#1{\mathbf{Y}_{\ell m}^{#1}}%
\global\long\def\sgsh#1#2{Y_{#2}^{#1}}%
\global\long\def\sgshlm#1{Y_{\ell m}^{#1}}%
\global\long\def\PB#1#2{\mathbf{P}_{#2}^{#1}}%
\global\long\def\PBjm#1{\mathbf{P}_{jm}^{#1}}%
\global\long\def\PBlm#1{\mathbf{P}_{\ell m}^{#1}}%
\global\long\def\Hansen#1#2{\mathbf{H}_{#2}^{#1}}%
\global\long\def\Hansenlm#1{\mathbf{H}_{\ell m}^{#1}}%
\global\long\def\gfnomega#1#2#3{G_{#2,#3\omega}^{#1}}%
 
\global\long\def\gfnlomega#1#2{G_{#2,\ell\omega}^{#1}}%

\title{Helioseismic finite-frequency sensitivity kernels for flows in spherical geometry including systematic effects}

\author[0000-0001-6433-6038]{Jishnu Bhattacharya}

\affil{Center for Space Science, New York University Abu Dhabi, PO Box -
129188, Abu Dhabi, United Arab Emirates}

\begin{abstract}
    Helioseismic inferences of large-scale flows in the solar interior necessitate accounting for the curvature of the Sun, both in interpreting systematic trends introduced in measurements as well as the sensitivity kernel that relates photospheric measurements to subsurface flow velocities. Additionally the inverse problem that relates measurements to model parameters needs to be well-posed to obtain accurate inferences, which necessitates a sparse set of parameters. Further, the sensitivity functions need to be computationally easy to evaluate. In this work we address these issues by demonstrating that the sensitivity kernels for flow velocities may be computed efficiently in a basis of vector spherical harmonics. We are also able to account for line-of-sight projections in Doppler measurements, as well as center-to-limb differences in line-formation heights. We show that given the assumed spherical symmetry of the background model, it is often cheap to simultaneously compute the kernels for pairs of observation points that are related by a rotation. Such an approach is therefore particularly well-suited to inverse problems for large-scale flows in the Sun, such as meridional circulation.
\end{abstract}

\section{Introduction}

Observations of the solar surface have revealed that the Sun harbors flows at a wide range of spatial and temporal scales \citep[see][for a review]{2005LRSP....2....6G}. These range from megameter-scaled convective cells referred to as granulation, to meridional circulations that span the expanse of the Sun and are believed to play an active part in angular momentum and flux transport processes \citep{2005LRSP....2....1M}. Flows at the largest scales also happen to be the ones that are understood poorly, and considerable efforts have been put into improving inference schemes in recent times to address this shortcoming in our understanding. Helioseismology enables us to relate surface measurements of seismic waves to convective flows within the Sun, therefore supplying us with an acoustic probe into the electromagnetically opaque solar interior. Seismic techniques have been applied to detect large-scale flows such as differential rotation \citep{1998ApJ...505..390S} and associated global features such as torsional oscillation and the near-surface shear layer, meridional circulation \citep{1996ApJ...460.1027H}, photospheric signatures of giant cells that are supposed to be associated with deep convection \citep{2013Sci...342.1217H}, although --- aside from rotational features --- a consensus on their properties has not yet been achieved. Meridional flows are of particular importance, as they are believed to play a key role in flux-transport dynamo models by conveying magnetic flux equator-wards at the bottom of the solar convection zone \citep{Dikpati_1999}, therefore understanding their subsurface profiles stands as an outstanding challenge for helioseismology. Presently their subsurface profiles are fairly uncertain, with \citet{2012ApJ...749L..13H} and \citet{2015ApJ...805..133J} suggesting a shallow return flow, \citet{2013ApJ...778L..38S,2013ApJ...774L..29Z} suggesting multiple cells in radius --- consistent with a shallow profile if the deeper cells remain undetected, while \citet{2015ApJ...813..114R} and \citet{2018ApJ...863...39M} find a single cell spanning the entire solar convection zone. The latitudinal profiles inferred also differ, with \citet{2013ApJ...778L..38S} suggesting multiple latitudinal cells, contrary to the other results. A careful study of the systematics involved in the analysis techniques might be necessary to unravel the differences in the conclusions reached by the various authors.

The standard solar model \citep[Model S,][]{1996Sci...272.1286C} is taken to be spherically symmetric, therefore seismic normal mode wavefunctions may be labelled by spherical harmonic degrees. Departures from spherical symmetry induced by convective flows lead to power being transferred between different wave modes, and a comparison of the deviation from a reference symmetric model would allow us to pinpoint the magnitude of the subsurface inhomogeneity. Various techniques have been used in the past to achieve this, ranging from time-distance helioseismology \citep{1993Natur.362..430D} that uses differences in seismic wave travel-times to estimate subsurface flows, ring-diagram analysis \citep{1988ApJ...333..996H} that uses shifts in the seismic power spectrum, to mode-couplings in Fourier space \citep{2007ApJ...668.1189W} that uses direct correlations between the wave modes for the estimation \citep[See][for a review]{2005LRSP....2....6G}. In this work we focus on time-distance helioseismology to frame an inverse problem and relate surface observations of seismic wave travel-times to subsurface flows.

An inference about the solar interior is usually drawn through an inverse problem that relates seismic wave parameters --- such as the travel-times of seismic waves --- to subsurface inhomogeneities, and the function that relates the two is referred to as the sensitivity kernel. This function encapsulates the physics of the solar model as well as the measurement procedure, and an accurate estimation of subsurface flows therefore requires a computation of the kernel that correctly accounts for the physics of wave propagation and the systematic effects associated with the measurement. A  second challenge that needs to be overcome is that of an ill-conditioned inverse problem, given that the number of parameters to infer often vastly outnumbers the measurements available. Such an inference may be aided by rephrasing the inverse problem in terms of an alternate, smaller set of parameters. Luckily such a set is readily available --- that of the reciprocal space, which --- in spherical geometry --- is spanned by spherical harmonics. Large scale flows on the Sun may be described in terms of a limited set of low-degree spherical harmonics. Additionally, improving the signal-to-noise ratio of the measured seismic parameters often involves careful averaging, which necessitates multiple evaluations of the kernel. The computation of the sensitivity kernel therefore needs to be computationally efficient as well. In this work we present an approach to compute sensitivity kernels that is able to address each of these issues.

The set of seismic eigenfunctions in the Sun forms a complete basis, therefore the kernel may be expanded in this basis and expressed as a sum of normal modes. This approach naturally incorporates the geometry of the Sun through the profile of the eigenfunctions. Initial attempts at computing finite-frequency sensitivity kernels had assumed a Cartesian background medium \citep{2007AN....328..228B,2007ApJ...671.1051J,2015SSRv..196..201B}, however large-scale flows sense the curvature of the Sun so the an analysis to infer them needs to be carried out by accounting for spherical geometry. Such an approach had been used by \cite{2016ApJ...824...49B} to compute kernels for seismic wave travel times derived from cross-covariances, and a variant was used by \cite{2017ApJ...842...89M} to compute kernels for travel times that were derived directly from wave velocities. \cite{Gizon2017A&A...600A..35G} proposed an alternate approach where the kernels are computed numerically assuming azimuthal symmetry. This approach that does not necessitate spherical symmetry, therefore it is more flexible than predecessors. All of these approaches are however computationally intensive, as was demonstrated by \cite{2018A&A...616A.156F}, where the authors explored an alternate approach: compute the spherical-harmonic coefficients of the kernel instead its spatial profile, and parameterize the inverse problem in terms of these coefficients. The work presented in our paper follows a similar approach. We show that it is possible to include line-of-sight projections and differences in line-formation heights into the modelled cross-covariances, thereby potentially alleviating systematic trends that exist in seismic measurements. Much of the fundamentals of the analysis technique were developed by \citet{2020ApJ...895..117B} in the context of subsurface sound-speed perturbations, and this work extends the analysis to flows. Finally, such an approach need not be confined to travel-time analysis. \citet{2017A&A...599A.111N} had demonstrated that it is straightforward to include amplitudes of seismic wave covariances to constrain the inverse problem, which --- used alongside travel times --- might lead to more accurate results.

\section{Vector Spherical Harmonics}

\subsection{Helicity basis}

The analysis of vector fields in spherical-polar coordinates is convenient in a basis that is a complex linear combination of the basis vectors $\mathbf{e}_r$, $\mathbf{e}_\theta$ and $\mathbf{e}_\phi$, given by

\begin{equation}
\begin{aligned}
\mathbf{e}_{+1} & =-\frac{1}{\sqrt{2}}\left(\mathbf{e}_{\theta}+i\mathbf{e}_{\phi}\right),\\
\mathbf{e}_{0} & =\mathbf{e}_{r}, \\
\mathbf{e}_{-1} & =\frac{1}{\sqrt{2}}\left(\mathbf{e}_{\theta}-i\mathbf{e}_{\phi}\right).
\end{aligned}
\label{eq:helicity_spherical}
\end{equation}

We follow \citet{1988qtam.book.....V} and refer to this basis as the ``helicity" basis.

\subsection{Definition of the harmonics}

Vector spherical harmonics (VSH) --- which are vector eigenfunctions of the Laplacian $\nabla^{2}$ on the unit sphere --- form a complete basis to expand vector fields in spherical geometry. We refer the readers to \citet{1988qtam.book.....V} and \citet{1976RSPTA.281..195J} for a detailed introduction to these functions, and we state the important results that we use in this work. We refer the readers to \cite{2020ApJ...895..117B} for an introduction to the specific functions used here. We use two linearly related bases in our analysis that may be defined at a point $\hat{n}=(\theta,\phi)$ in terms of the spherical harmonic $Y_{\ell m}\left(\hat{n}\right)$ as:
\begin{enumerate}
\item Hansen VSH \citep{PhysRev.47.139,1957ApJ...126..457C}, defined as
\begin{align}
\Hansenlm{\left(-1\right)}\left(\hat{n}\right) & =Y_{\ell m}\left(\hat{n}\right)\mathbf{e}_{r},\nonumber \\
\Hansenlm{\left(0\right)}\left(\hat{n}\right) & =\frac{-i}{\sqrt{\ell\left(\ell+1\right)}}\mathbf{e}_{r}\times\grad_{\Omega}Y_{\ell m}\left(\hat{n}\right),\\
\Hansenlm{\left(1\right)}\left(\hat{n}\right) & =\frac{1}{\sqrt{\ell\left(\ell+1\right)}}\grad_{\Omega}Y_{\ell m}\left(\hat{n}\right).\nonumber 
\end{align}
\item Phinney-Burridge (PB) VSH \citep{1973GeoJ...34..451P}, that may be expressed as a linear combination of the Hansen VSH basis as
\begin{equation}
\begin{aligned}\PBlm{+1} & =\frac{1}{\sqrt{2}}\left(\Hansenlm{\left(1\right)}-\Hansenlm{\left(0\right)}\right),\\
\PBlm{-1} & =\frac{1}{\sqrt{2}}\left(\Hansenlm{\left(1\right)}+\Hansenlm{\left(0\right)}\right),\\
\PBlm 0 & =\Hansenlm{\left(-1\right)},
\end{aligned}
\label{eq:PB_Hansen_conversion}
\end{equation}
The two bases are related through a rotation by $\pi/4$ about $\mathbf{e}_r$.

\end{enumerate}
The analysis scheme hinges on the fact that the Green functions are expressed most easily in the Hansen basis, whereas their components
in the spherical-polar basis are easier to represent in the PB basis. The contravariant components of the PB basis vectors $\PBlm{\gamma}$ in the helicity
basis are 
\begin{equation}
\left[\PBlm{\gamma}\right]^\alpha  =\sqrt{\frac{2\ell+1}{4\pi}}\,d_{m\alpha}^{\ell}\left(\theta\right)e^{im\phi} \delta_{\alpha,\gamma}, 
\end{equation}
where $d_{m\alpha}^{\ell}\left(\theta\right)$ is an element of the Wigner d-matrix and $\delta_{\alpha,\gamma}$ is the Kronecker delta function. We follow \cite{DahlenTromp} and refer to the diagonal components $\left[\PBlm{\gamma}\right]^\gamma$ as generalized spherical harmonics, defined as 
\begin{equation}
\sgshlm{\gamma} \left(\theta,\phi\right) = \sqrt{\frac{2\ell+1}{4\pi}} d_{m\gamma}^{\ell}\left(\theta\right)e^{im\phi}.
\end{equation}

The fact that the PB VSH are diagonal in the helicity basis plays a pivotal role in the analysis presented in this work, and allows seamless conversions between a basis of VSH and the spherical-polar one.

\subsection{Derivatives of vector spherical harmonics}

The derivative of the VSH may be computed in the PB VSH following the relations described by \citet{1973GeoJ...34..451P} and \citet{DahlenTromp}, and we choose to retain the notation used by the latter. We may expand a function $f\left(\mathbf{x}\right)$ in the PB VSH basis as 
\[
\mathbf{f}\left(\mathbf{x}\right)=\sum_{\ell m\alpha}f_{\ell m}^{\alpha}\left(r\right)\PB{\alpha}{\ell m}\left(\hat{n}\right).
\]
The gradient of $\mathbf{f}\left(\mathbf{x}\right)$ may be expressed as a sum over the gradients of the components. In the helicity basis,
we obtain 
\begin{align}
\grad\left[f_{\ell m}^{\alpha}\left(r\right)\PBlm{\alpha}\left(\hat{n}\right)\right] & =\left(\frac{d}{dr}f_{\ell m}^{\alpha}\left(r\right)\right)\mathbf{e}_{0}\PBlm{\alpha}\left(\hat{n}\right)+\nonumber \\
 & \frac{1}{r}f_{\ell m}^{\alpha}\left(r\right)\left[\Omega_{\ell}^{\alpha}\sgshlm{-1+\alpha}\left(\hat{n}\right)\mathbf{e}_{-1}\mathbf{e}_{\alpha}-\sgshlm{\alpha}\left(\hat{n}\right)\mathbf{e}_{-1}\mathbf{e}_{\alpha+1}\right.\nonumber \\
 & \left.+\Omega_{\ell}^{-\alpha}\sgshlm{1+\alpha}\left(\hat{n}\right)\mathbf{e}_{+1}\mathbf{e}_{\alpha}-\sgshlm{\alpha}\left(\hat{n}\right)\mathbf{e}_{+1}\mathbf{e}_{\alpha-1}\right],\label{eq:grad_VSH_PB}
\end{align}
where $\mathbf{e}_{\alpha}=0$ for $\left|\alpha\right|>1$, and $\Omega_{\ell}^{\alpha}=\sqrt{\left(\ell+\alpha\right)\left(\ell-\alpha+1\right)/2}$.

\subsection{Integral of the three-term product}

One of the key steps in the analysis is evaluating the angular integral
\begin{equation}
I_{\ell_{1}m_{1}\ell_{2}m_{2}\ell_{3}m_{3}}^{n_{1}n_{2}n_{3}}\left(f_{\ell_{3}}^{n_{3}}\left(r\right)\right)=\int d\hat{n}\PB{n_{1}}{\ell_{1}m_{1}}\left(\hat{n}\right)\cdot\left[\PB{n_{2}}{\ell_{2}m_{2}}\left(\hat{n}\right)\cdot\grad\left(f_{\ell_{3}}^{n_{3}}\left(r\right)\PB{n_{3}}{\ell_{3}m_{3}}\left(\hat{n}\right)\right)\right].
\end{equation}
We evaluate the integral in Appendix \ref{sec:Appendix_VSH-triple-integral}, and show that it may be expressed in the form 
\begin{align}
I_{\ell_{1}m_{1}\ell_{2}m_{2}\ell_{3}m_{3}}^{n_{1}n_{2}n_{3}}\left(f_{\ell_{3}}^{n_{3}}\left(r\right)\right) & =\left(-1\right)^{m_{2}}C_{\ell_{1}m_{1}\ell_{3}m_{3}}^{\ell_{2}-m_{2}}J_{\ell_{1}\ell_{2}\ell_{3}}^{n_{2}n_{1}n_{3}}\left(f_{\ell_{3}}^{n_{3}}\left(r\right)\right),\label{eq:triple_int}
\end{align}
where
\begin{align}
J_{\ell_{2}\ell_{1}\ell_{3}}^{n_{2}n_{1}n_{3}}\left(f\left(r\right)\right) & =\eta_{\ell_{2}}^{\ell_{1}\ell_{3}}\left(-1\right)^{n_{3}}\left[\delta_{n_{2},0}\left(\frac{d}{dr}f\left(r\right)\right)C_{\ell_{1}-n_{3}\ell_{3}n_{3}}^{\ell_{2}0}\delta_{n_{1},-n_{3}}\right.\nonumber \\
 & +\delta_{n_{2},1}\frac{1}{r}f\left(r\right)\left\{ \Omega_{\ell}^{n_{3}}C_{\ell_{1}-n_{3}\ell_{3}n_{3}-1}^{\ell_{2}-1}\delta_{n_{1},-n_{3}}+C_{\ell_{1}-n_{3}-1\ell_{3}n_{3}}^{\ell_{2}-1}\delta_{n_{1},-n_{3}-1}\delta_{n_{3}}^{-1,0}\right\} \nonumber \\
 & \left.+\delta_{n_{2},-1}\frac{1}{r}f\left(r\right)\left\{ \Omega_{\ell}^{-n_{3}}C_{\ell_{1}-n_{3}\ell_{3}n_{3}+1}^{\ell_{2}1}\delta_{n_{1},-n_{3}}+C_{\ell_{1}-n_{3}+1\ell_{3}n_{3}}^{\ell_{2}1}\delta_{n_{1},-n_{3}+1}\delta_{n_{3}}^{0,1}\right\} \right],\label{eq:J}\\
\eta_{\ell_{2}}^{\ell_{1}\ell_{3}} & =\sqrt{\frac{\left(2\ell_{1}+1\right)\left(2\ell_{3}+1\right)}{4\pi\left(2\ell_{2}+1\right)}},
\end{align}
$C_{\ell_{1}m_{1}\ell_{3}m_{3}}^{\ell_{2}-m_{2}}$ is the Clebsch-Gordan coefficient that connects the sum of the angular momenta $(\ell_1, m_1)$ and $(\ell_3, m_3)$ to $(\ell_2, -m_2)$, and $\delta_{a}^{b,c}=\delta_{a,b}+\delta_{a,c}$ is the sum of two Kronecker delta functions. The function $J_{\ell_{2}\ell_{1}\ell_{3}}^{n_{2}n_{1}n_{3}}\left(f\left(r\right)\right)$ satisfies the symmetry relation
\begin{equation}
J_{\ell_{2}\ell_{1}\ell_{3}}^{-n_{2}-n_{1}-n_{3}}\left(f\left(r\right)\right)=\left(-1\right)^{\ell+j_{1}+j_{2}}J_{\ell_{2}\ell_{1}\ell_{3}}^{n_{2}n_{1}n_{3}}\left(f\left(r\right)\right).\label{eq:J_symmetry}
\end{equation}
We note that the $J_{\ell_{2}\ell_{1}\ell_{3}}^{n_{2}n_{1}n_{3}}\left(f\left(r\right)\right)$ is non-zero only for the values of $\ell_{1}$, $\ell_{2}$ and $\ell_{3}$ that satisfy the triangle inequality $\left|\ell_{1}-\ell_{2}\right|\leq\ell_{3}\leq\ell_{1}+\ell_{2}$.

\section{Seismic measurements on the Sun}
Acoustic waves in the Sun are excited by vigorous transonic, non-adiabatic convective flows near the photosphere, and these waves subsequently traverse the solar interior to re-emerge and be detected at the surface of the Sun. A key seismic measurement is that of the line-of-sight projected wave velocity inferred from Doppler shifts of atmospheric absorption lines in the Sun. We choose to work in temporal-frequency domain bearing in mind that the background medium is temporally stationary. We work in spherical polar coordinates with the origin at the center of the Sun. A point $\mathbf{x}$ in the Sun may be described by its radial coordinate $r$, its co-latitude $\theta$ and azimuth $\phi$. We also use the notation $\hat{n}=(\theta,\phi)$ to denote a point on a shell at a fixed radius $r$. The isotropic background solar model at equilibrium may be described in terms of the radial profiles of the density $\rho$, pressure $p$, gravitational acceleration $\mathbf{g}$ and sound-speed $c$. The equation governing the propagation of seismic waves in temporal frequency domain at a point $\mathbf{x}=(r,\theta,\phi)$ in the Sun, given a source distribution $\mathbf{S}(\mathbf{x},\omega)$, may be represented in terms of the wave displacement $\bm{\xi}(\mathbf{x},\omega)$ as

\begin{equation}
-\rho\omega^{2}\bm{\xi}-2i\omega\gamma\bm{\xi}  - \grad\left(\rho c^{2}\grad\cdot\bm{\xi}-\rho\bm{\xi}\cdot\mathbf{e}_{r}g\right)-g\mathbf{e}_{r}\grad\cdot\left(\rho\bm{\xi}\right) = \mathbf{S}(\mathbf{x},\omega),
\label{eq:waveeqn}
\end{equation}
where the frequency-dependent constant $\gamma$ denotes the attenuation experienced by the wave, and we have suppressed the coordinate-dependence on the left-hand side to simplify the notation. We follow the approach of \citep{2020ApJ...895..117B} and consider the damping constant $\gamma$ to be a polynomial function of the temporal frequency.
We condense the notation by referring to the terms on the left-hand side of Equation \eqref{eq:waveeqn} collectively as $\mathcal{L}\bm{\xi}(\mathbf{x},\omega)$, where the frequency-dependent wave operator $\mathcal{L}$ incorporates the spatial derivatives.

Doppler measurements of seismic waves on the Sun are sensitive to the line-of-sight projected component of the velocity. We assume that seismic observations are carried out at a point $\mathbf{x}_\mathrm{obs}=(r_\mathrm{obs},\theta_\mathrm{obs},\phi_\mathrm{obs})$. This is a great simplification of the actual process of line-formation, since spectral lines form over a broad range of heights in an unsteady atmosphere, therefore observations are not limited to a specific spatial location. We may interpret the radial coordinate $r_\mathrm{obs}$ as an average line-formation height, which is around $150$ km above the photosphere at the disk center for the Fe $6173\,\mbox{\normalfont\AA}$ line \citep{2011SoPh..271...27F} that the Helioseismic and Magnetic Imager \citep[HMI,][]{2012SoPh..275..207S} is sensitive to.

The line-of-sight projected velocity may be expressed in the frequency domain in terms of the line-of-sight vector $\bm{l}\left(\mathbf{x}_\mathrm{obs}\right)$ and the wave displacement $\bm{\xi}\left(\mathbf{x}_\mathrm{obs},\omega\right)$ as 
\begin{equation}
v\left(\mathbf{x}_\mathrm{obs},\omega\right)=i\omega\,\bm{l}\left(\mathbf{x}_\mathrm{obs}\right)\cdot\bm{\xi}\left(\mathbf{x}_\mathrm{obs},\omega\right).
\label{eq:v_Doppler}
\end{equation}
The radial coordinate $r_\mathrm{obs}$ that an observation is sensitive to depends on the angular distance of the observation point from the disk center \citep{2015ApJ...808...59K}, which introduces a weak angular dependence on $r_\mathrm{obs}$. We note that the actual measured value will be a convolution of the projected velocity with the point-spread function of the detector, however we do not consider this in the present analysis.

The position-dependence of the line-of-sight vector $\bm{l}\left(\mathbf{x}\right)$ is weak owing to the fact that the distance between the Sun and the Earth is significantly larger than the solar radius $\left(R_{\odot}\approx0.0046\,\mathrm{AU}\right)$, so in practice the line-of-sight direction might be assumed to be identical at all points on the Sun without incurring significant errors. We retain the dependence in subsequent analysis as it does not pose any additional algebraic challenge. Despite the notation used in this work, the line-of-sight vector actually depends on two spatial points -- the point $\mathbf{x}$ on the Sun where seismic wave velocities are measured, as well as the spatial location of the detector. This implies that if we change only the measurement point keeping the detector location fixed, the line-of-sight direction does not transform as a vector field. This issue, however, does not pose a challenge to us as the vector may be trivially recomputed at each measurement point.

Waves on the Sun are excited stochastically by near-surface convection, and the wave sources may be modeled as a Gaussian random process. We follow \citet{2016ApJ...824...49B} and assume that the wave sources are purely radial. This is a simplifying assumption motivated by the fact that the highest flow velocities at the surface are detected in granular downdrafts, however our analysis does not depend fundamentally on this assumption. We denote the source distribution by $\mathbf{S}\left(\mathbf{x},\omega\right)=S_{r}\left(\mathbf{x},\omega\right)\mathbf{e}_{r}$, where the radial component $S_{r}$ has a mean of zero, and a covariance that may be modeled to be isotropic and limited to a shell of radius $r_{\mathrm{src}}$:
\begin{equation}
\left\langle S_{r}^{*}\left(\mathbf{x}_{1};\omega\right)S_{r}\left(\mathbf{x}_{2};\omega\right)\right\rangle =P\left(\omega\right)\delta\left(\mathbf{x}_{1}-\mathbf{x}_{2}\right)\frac{1}{r_{\mathrm{src}}^{2}}\delta\left(\left|\mathbf{x}_{1}\right|-r_{\mathrm{src}}\right),
\label{eq:S_cov}
\end{equation}
where $P(\omega)$ represents the frequency dependence of the source covariance, and the angular brackets denote an ensemble average. We assume $P(\omega)$ to be a Gaussian in this work with a mean of $\omega_0 = 2\pi\times 3\,\mathrm{mHz}$ and a width of $\Delta\omega = 2\pi\times 0.4\,\mathrm{mHz}$. The amplitude of $P(\omega)$ has been arbitrarily chosen to be $1$ as this does not affect travel-time measurements, however this needs to be calibrated for a full-waveform inversion. We choose the source to be located at $75$ km below the photosphere.
This model of the source covariance is inspired by simulations such as those by \citet{1991LNP...388..141N}, where it has been demonstrated that the excitation of waves take place in regions of high non-adiabatic pressure as well as turbulent pressure fluctuations, which occur in the Sun in a thin layer of width around a hundred kilometers below the photosphere. A more realistic model might include a radial profile of the source covariance, however this would significantly increase the computational cost and is beyond the scope of the present work.

A Gaussian source also implies that the wave displacement is a zero-mean Gaussian random variable. The fundamental measurement that interests us therefore is the two-point covariance of seismic waves $C(\mathbf{x}_1,\mathbf{x}_2,\omega)$ \citep{1993Natur.362..430D}. A change in the solar model affects the propagation of seismic waves in the Sun, and consequently alters the measured cross-covariance. In the following sections we develop the formalism to relate changes in the solar model to that of seismic wave travel-times projected from the cross-covariance, focusing specifically on changes introduced by flows in the solar interior.

\subsection{Green function}

Propagation of seismic waves in the Sun is governed by Equation \eqref{eq:waveeqn}, which may be rewritten in terms of the Green function $\mathbf{G}\left(\mathbf{x}_{\mathrm{obs}},\mathbf{x}_{\mathrm{src}},\omega\right)$ that describes the impulse response of the wave equation given an excitation at $\mathbf{x}_\mathrm{src}$ and a measurement at $\mathbf{x}_\mathrm{obs}$. The wave displacement $\bm{\xi}(\mathbf{x},\omega)$ is related to the sources $\mathbf{S}(\mathbf{x},\omega)$ through the Green function as 
\begin{align}
\bm{\xi}\left(\mathbf{x}_{\mathrm{obs}},\omega\right) & =\int d\mathbf{x}_{\mathrm{src}}\,\mathbf{G}\left(\mathbf{x}_{\mathrm{obs}},\mathbf{x}_{\mathrm{src}},\omega\right)\cdot
\mathbf{S}\left(\mathbf{x}_{\mathrm{src}},\omega\right).\label{eq:xi_G_S}
\end{align}
We refer the readers to \cite{2020ApJ...895..117B} where the authors had described the numerical computation of the Green function. We may expand the Green function in the PB VSH basis as 
\begin{equation}
\mathbf{G}\left(\mathbf{x}_{\mathrm{obs}},\mathbf{x}_{\mathrm{src}},\omega\right)=\sum_{\alpha,\beta=\pm1}\sum_{jm}\gfnomega{\alpha}{\beta}j\left(r_{\mathrm{obs}},r_{\mathrm{src}}\right)\PBjm{\alpha}\left(\hat{n}_{\mathrm{obs}}\right)\PB{\beta*}{jm}\left(\hat{n}_{\mathrm{src}}\right).\label{eq:G_PB_VSH}
\end{equation}
The components of the Green function satisfy the symmetry relations $\gfnomega{\pm\alpha}{\pm\beta}j=\gfnomega{\alpha}{\beta}j$ owing to the fact that the seismic eigenfunctions in the Sun lack a toroidal component. The Green tensor therefore has four independent components, and without loss of generality we choose these to be $G_{0}^{0}$, $G_{0}^{1}$, $G_{1}^{0}$ and $G_{1}^{1}$.

The Green function satisfies the reciprocity relation $\mathbf{G}\left(\mathbf{x}_{\mathrm{obs}},\mathbf{x}_{\mathrm{src}},\omega\right)=\mathbf{G}^{T}\left(\mathbf{x}_{\mathrm{src}},\mathbf{x}_{\mathrm{obs}},\omega\right)$, which may be expressed in terms of the components as 
\begin{equation}
G_{\beta}^{\alpha}\left(r_{\mathrm{src}},r_{\mathrm{obs}},\omega\right)=G_{\alpha}^{\beta}\left(r_{\mathrm{obs}},r_{\mathrm{src}},\omega\right).\label{eq:reciprocity}
\end{equation}We denote the components of the Green function corresponding to a radial source by the symbol $\mathbf{G}_{r}$, which is defined by restricting Equation (\ref{eq:G_PB_VSH}) to $\beta=0$. We obtain 
\begin{equation}
\mathbf{G}_{r}\left(\mathbf{x}_{\mathrm{obs}},\mathbf{x}_{\mathrm{src}},\omega\right)=\sum_{\alpha=\pm1}\sum_{jm}\gfnomega{\alpha}0j\left(r_{\mathrm{obs}},r_{\mathrm{src}}\right)\PBjm{\alpha}\left(\hat{n}_{\mathrm{obs}}\right)Y_{jm}^{*}\left(\hat{n}_{\mathrm{src}}\right).
\end{equation}

We compute the radial profiles of $G^\alpha_{\beta,j\omega}(r,r_\mathrm{src})$ numerically using a finite-difference scheme following \citet{2020ApJ...895..117B}.

\subsection{Cross-covariance}

\begin{figure}
    \includegraphics[scale=0.9]{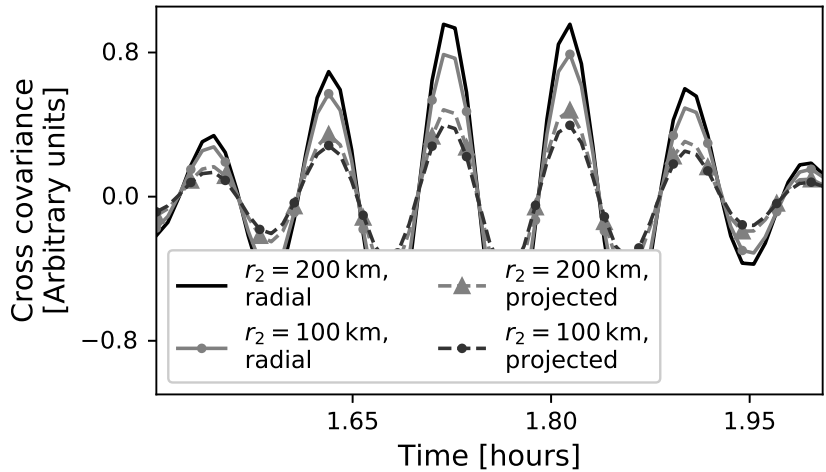}
    \caption{Cross-covariance as a function of time measured between the observation points $\mathbf{x}_1=(R_\odot+200\,\mathrm{km},\pi/2,0)$ and $\mathbf{x}_2=(R_\odot+r_2,\pi/2,\pi/3)$. The legend indicates the height $r$ above the photosphere (in km) at which the observation are carried out for the point $\mathbf{x}_2$. The solid lines represent covariances between the radial components of the wave velocities, whereas the dashed lines represent covariance between line-of-sight projected wave velocities.}
    \label{fig:Ctlosheight}
\end{figure}

The line-of-sight projected velocity $v(\mathbf{x}_\mathrm{obs},\omega)$ from Equation \eqref{eq:v_Doppler} is usually modelled as a zero-mean random variable, so its covariance represents the fundamental measurement in time-distance seismology. The covariance of the Doppler signal may be expressed in terms of the wave displacement as 
\begin{align}
C\left(\mathbf{x}_{1},\mathbf{x}_{2},\omega\right) & =\left\langle v^{*}\left(\mathbf{x}_{1},\omega\right) \,v\left(\mathbf{x}_{2},\omega\right)\right\rangle =\omega^{2}\left\langle \bm{l}\left(\mathbf{x}_{1}\right)\cdot\bm{\xi}^{*}\left(\mathbf{x}_{1},\omega\right)\,\bm{l}\left(\mathbf{x}_{2}\right)\cdot\bm{\xi}\left(\mathbf{x}_{2},\omega\right)\right\rangle .
\end{align}
Using Equation (\ref{eq:xi_G_S}) and our model for the source covariance from Equation (\ref{eq:S_cov}), we may express the covariance in terms of the Green function as 
\begin{equation}
C\left(\mathbf{x}_{1},\mathbf{x}_{2},\omega\right)=\omega^{2}P\left(\omega\right)\int d\Omega_{\mathrm{src}}\,\bm{l}\left(\mathbf{x}_{1}\right)\cdot\mathbf{G}_{r}^{*}\left(\mathbf{x}_{1},\mathbf{x}_{\mathrm{src}};\omega\right)\bm{l}\left(\mathbf{x}_{2}\right)\cdot\mathbf{G}_{r}\left(\mathbf{x}_{2},\mathbf{x}_{\mathrm{src}};\omega\right),
\end{equation}
where the integral is carried out over the angular distribution of the sources. We may evaluate the angular part of this integral analytically using the separation of variables of the Green function in the PB VSH basis (Equation \ref{eq:G_PB_VSH}), to obtain
\begin{equation}
C\left(\mathbf{x}_{1},\mathbf{x}_{2},\omega\right)=\omega^{2}P\left(\omega\right)\sum_{\alpha,\beta=-1}^{1}\sum_{jm}\gfnomega{\alpha*}0j\left(r_{1},r_{\mathrm{src}}\right)\gfnomega{\beta}0j\left(r_{2},r_{\mathrm{src}}\right)\bm{l}\left(\mathbf{x}_{1}\right)\cdot\PBjm{\alpha*}\left(\hat{n}_{1}\right)\bm{l}\left(\mathbf{x}_{2}\right)\cdot\PBjm{\beta}\left(\hat{n}_{2}\right).
\label{eq:Comegaexpr}
\end{equation}
We may recast the expression as 
\begin{equation}
C\left(\mathbf{x}_{1},\mathbf{x}_{2},\omega\right)=\bm{l}\left(\mathbf{x}_{1}\right)\bm{l}\left(\mathbf{x}_{2}\right):\mathbf{C}\left(\mathbf{x}_{1},\mathbf{x}_{2},\omega\right),
\label{eq:Ctensor}
\end{equation}
where the $3\times3$ rank-$2$ tensor 
\begin{equation}
    \mathbf{C}\left(\mathbf{x}_{1},\mathbf{x}_{2},\omega\right) = \omega^{2}P\left(\omega\right)\sum_{\alpha,\beta=-1}^{1}\sum_{jm}\gfnomega{\alpha*}0j\left(r_{1},r_{\mathrm{src}}\right)\gfnomega{\beta}0j\left(r_{2},r_{\mathrm{src}}\right)\PBjm{\alpha*}\left(\hat{n}_{1}\right)\PBjm{\beta}\left(\hat{n}_{2}\right)
\end{equation} 
captures the covariance between the various components of the velocity of seismic waves, and the colon indicates a double contraction $\left(\mathbf{A}:\mathbf{B}=A_{ij}B_{ij}\right)$. We plot the cross-covariance as a function of time in Figure \ref{fig:Ctlosheight} for two different combinations of observation heights, and by including as well as ignoring line-of-sight projection. We show that the results are sensitive to the systematic effects chosen, therefore precise modelling of the cross-covariances might need to account for these.

\begin{figure}
    \includegraphics[scale=0.8]{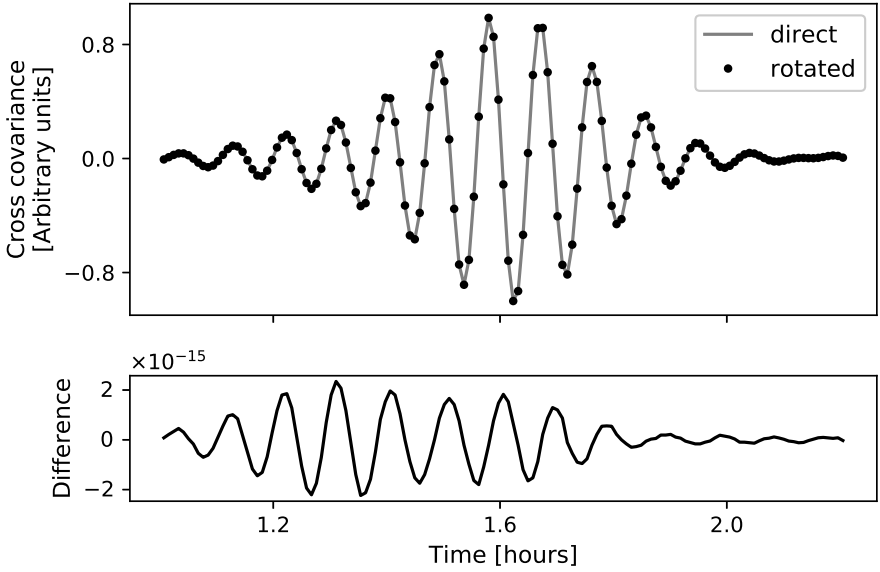}
    \caption{Line-of-sight projected cross-covariance as a function of time. The solid line is computed directly using Equation \eqref{eq:Comegaexpr} for $\mathbf{x}_1=(R_\odot+200\,\mathrm{km},\pi/2,\pi/12)$ and $\mathbf{x}_2=(R_\odot+200\,\mathrm{km},\pi/2,\pi/3)$, and the dots are computed for $\mathbf{x}^\prime_1=(R_\odot+200\,\mathrm{km},\pi/2,\pi/6)$ and $\mathbf{x}^\prime_2=(R_\odot+200\,\mathrm{km},\pi/2,5\pi/12)$ followed by rotating the coordinate system by $\pi/12$ about $\mathbf{e}_z$ to align $\mathbf{x}^\prime_i$ with $\mathbf{x}_i$. The close match demonstrates the transformation of the cross-covariance as a scalar under rotation.}
    \label{fig:Crot}
\end{figure}

The advantage of rewriting the expression in the form as in Equation \eqref{eq:Ctensor} is that the tensor $\mathbf{C}$ is a function only of the measurement points $\mathbf{x}_1$ and $\mathbf{x}_2$, and does not depend on the detector location. This also means that under rotation of the observation points on the surface of an isotropic model of the Sun, the covariance $\mathbf{C}$ transforms as a scalar, that is $\left[\mathbf{C}(\mathbf{x}_1^\prime,\mathbf{x}_2^\prime,\omega)\right]^{\alpha \beta} = \left[\mathbf{C}(\mathbf{x}_1,\mathbf{x}_2,\omega)\right]^{\alpha \beta}$ where $\mathbf{x}_i^\prime$ is related to $\mathbf{x}_i$ through a rotation. The projection operator may be thought of as a final step carried out following the modeling of the covariance tensor of seismic wave velocities in the Sun. We demonstrate this rotational symmetry in Figure \ref{fig:Crot} for the points $\mathbf{x}_1\,=\,(R_\odot+200\,\mathrm{km},\pi/2,\pi/12)$ and $\mathbf{x}_2\,=\,(R_\odot+200\,\mathrm{km},\pi/2,\pi/3)$, where we compute the line-of-sight projected cross-covariance in two ways: (1) by using Equation \eqref{eq:Comegaexpr} directly for $\mathbf{x}_1$ and $\mathbf{x}_2$, and (2) by computing the tensor $\mathbf{C}(\mathbf{x}_1^\prime,\mathbf{x}_2^\prime,\omega)$ for $\mathbf{x}_1^\prime=(\pi/2,\pi/6)$ and $\mathbf{x}_2^\prime=(\pi/2,5\pi/12)$, and using the fact that it transforms as a scalar under rotation. We find a close match with the difference being almost entirely numerical, proving the ease of transforming tensors between pairs of points on the sphere that are related by a rotation. We note that such a rotational transformation crucially assumes a separation between the observation height and the angular coordinates, therefore this might not lead to accurate results if the angle of rotation is large, and the center-to-limb difference in line-formation height is significant.

\section{Flows as a perturbation}

Model S \citep{1996Sci...272.1286C} --- which is often used as a standard solar model --- is spherically symmetric and does not explicitly account for the advection of seismic waves by flows present in the Sun. Weak flows in the Sun are therefore treated as perturbations about this model, and their magnitudes and profiles may be inferred in the first Born approximation \citep{2002ApJ...571..966G}. We denote the flow velocity at a point $\mathbf{x}$ within the Sun by the symbol $\mathbf{u}\left(\mathbf{x}\right)$. The advection of seismic waves by the underlying velocity fields is represented by the operator $\delta\mathcal{L}\left(\mathbf{x};\omega\right)=2i\omega\rho\mathbf{u}\left(\mathbf{x}\right)\cdot\grad$ to a linear order in the flow velocity. The resulting advection due to flows alters the local wave speed, and changes the measured seismic signal at the solar surface.

We expand the velocity field in the PB VSH basis as 
\begin{equation}
\mathbf{u}\left(\mathbf{x}\right)=u_{00}^{0}\left(r\right)\PB 0{00}\left(\hat{n}\right)+\sum_{\gamma=-1}^{1}\sum_{\ell=1}^{\infty}\sum_{m=-\ell}^{\ell}u_{\ell m}^{\gamma}\left(r\right)\PBlm{\gamma}\left(\hat{n}\right).\label{eq:u_PB_VSH}
\end{equation}
The first term is purely radial and spherically symmetric, and we may choose to drop the term depending on the type of flow that we
are interested in. We use the shorthand $\sum_{\ell m\gamma}$ to denote $\sum_{\ell=0}^{\infty}\sum_{m=-\ell}^{\ell}\sum_{\gamma=-\min\left(1,\ell\right)}^{\min\left(1,\ell\right)}$ --- where $\min\left(1,\ell\right)$ chooses the minimum of $1$ and $\ell$ and restricts $\gamma$ to $0$ for $\ell=0$ --- and rewrite the flow field as 
\begin{equation}
\mathbf{u}\left(\mathbf{x}\right)=\sum_{\ell m\gamma}u_{\ell m}^{\gamma}\left(r\right)\PBlm{\gamma}\left(\hat{n}\right).\label{eq:u_PB_VSH_aux}
\end{equation}

\subsection{A change in the Green function}

The Green function $\mathbf{G}\left(\mathbf{x}_{i},\mathbf{x}_{\mathrm{src}};\omega\right) $ dictates the propagation of seismic waves having a frequency $\nu=\omega/2\pi$ between the points $\mathbf{x}_{src}$ and $\mathbf{x}_i$ in the Sun. A shift in wave propagation properties may therefore be described in terms of an altered Green function, one that differs from the original by $\delta\mathbf{G}\left(\mathbf{x}_{i},\mathbf{x}_{\mathrm{src}};\omega\right) $. Our goal is to connect a variation in wave propagation to  a corresponding difference in the background model of the Sun. A change in the wave operator by $\delta\mathcal{L}\left(\mathbf{x};\omega\right)$ leads to a variation in the Green function that may be computed in the first Born approximation to be 
\begin{align}
\delta\mathbf{G}\left(\mathbf{x}_{i},\mathbf{x}_{\mathrm{src}};\omega\right) & =-\int d\mathbf{x}\mathbf{G}\left(\mathbf{x}_{i},\mathbf{x};\omega\right)\cdot\left[\delta\mathcal{L}\left(\mathbf{x};\omega\right)\mathbf{G}\left(\mathbf{x},\mathbf{x}_{\mathrm{src}};\omega\right)\right],\label{eq:first_born_approx}
\end{align}
The integral is carried out over all the scattering points in the solar interior. We evaluate the angular part of the integral analytically using Equation (\ref{eq:triple_int}), and cast Equation (\ref{eq:first_born_approx}) in the form 
\begin{align}
\delta\mathbf{G}\left(\mathbf{x}_{i},\mathbf{x}_{\mathrm{src}};\omega\right) & =\sum_{\ell m\gamma}\int r^{2}dr\,u_{\ell m}^{\gamma}\left(r\right)\sum_{j_{1}j_{2}}\sum_{\alpha_{1}\beta_{2}}J_{\ell j_{1}j_{2}\omega;\alpha_{1}\beta_{2}}^{-\gamma}\left(r,r_{i},r_{\mathrm{src}}\right)\PB{j_{1}j_{2}\alpha_{1}\beta_{2}}{\ell m}\left(\hat{n}_{i},\hat{n}_{\mathrm{src}}\right),\label{eq:dG_PB_VSH}
\end{align}
where 
\begin{equation}
J_{\ell j_{1}j_{2}\omega;\alpha_{1}\beta_{2}}^{\gamma}\left(r,r_{i},r_{\mathrm{src}}\right)=-2i\omega\rho\,\sum_{\alpha_{2}\beta_{1}}\gfnomega{\beta_{1}}{\alpha_{1}}{j_{1}}\left(r,r_{i}\right)J_{\ell j_{1}j_{2}}^{\gamma\beta_{1}\alpha_{2}}\left(\gfnomega{\alpha_{2}}{\beta_{2}}{j_{2}}\left(r,r_{\mathrm{src}}\right)\right),\label{eq:Jgamma_defn}
\end{equation}
with $J_{\ell j_{1}j_{2}}^{\gamma\beta_{1}\alpha_{2}}$ as defined in Equation (\ref{eq:J}), and the angular function $\PB{j_{1}j_{2}\alpha_{1}\beta_{2}}{\ell m}\left(\hat{n}_{i},\hat{n}_{\mathrm{src}}\right)$ is a bipolar spherical harmonic that couples the angular momenta $j_{1}$ and $j_{2}$ with $\ell$, defined as 
\begin{equation}
\PB{j_{1}j_{2}\alpha_{1}\beta_{2}}{\ell m}\left(\hat{n}_{i},\hat{n}_{\mathrm{src}}\right)=\sum_{m_{1}m_{2}}C_{j_{1}m_{1}j_{2}m_{2}}^{\ell m}\PB{\alpha_{1}}{j_{1}m_{1}}\left(\hat{n}_{i}\right)\PB{\beta_{2}}{j_{2}m_{2}}\left(\hat{n}_{\mathrm{src}}\right).
\end{equation}
We derive the relation in Appendix \ref{subsec:Appendix_Green-function}.  The radial component $J_{\ell j_{1}j_{2}\omega;\alpha_{1}\beta_{2}}^{\gamma}$ satisfies the following symmetry relations:

\begin{equation}
\begin{aligned}J_{\ell j_{1}j_{2}\omega;\alpha_{1}\beta_{2}}^{-\gamma} & =\left(-1\right)^{\ell+j_{1}+j_{2}}J_{\ell j_{1}j_{2}\omega;\alpha_{1}\beta_{2}}^{\gamma},\\
J_{\ell j_{1}j_{2}\omega;\pm\alpha_{1},\pm\beta_{2}}^{\gamma} & =J_{\ell j_{1}j_{2}\omega;\alpha_{1}\beta_{2}}^{\gamma}.
\end{aligned}
\label{eq:Jgamma_symemtry}
\end{equation}
The first equation tells us that $J_{\ell j_{1}j_{2}\omega;\alpha_{1}\beta_{2}}^{0}$ is non-zero only if $\ell+j_{1}+j_{2}$ is even.

Under the radial-source assumption, we only need to evaluate the terms $J_{\ell j_{1}j_{2}\omega;\alpha_{1}0}^{\gamma}$ for $\gamma=0$ and $\gamma=1$, bearing in mind that the $\gamma=-1$ term is related to the $\gamma=1$ term through Equation (\ref{eq:Jgamma_symemtry}).
We define the terms
\begin{align}
N_{\ell}^{j_{1}j_{2}} & =\sqrt{\frac{\left(2j_{1}+1\right)\left(2j_{2}+1\right)}{4\pi\left(2\ell+1\right)}},\\
\zeta_{\ell}^{j_{1}j_{2}} & =\frac{\left(\left(\Omega_{j_{1}}^{0}\right)^{2}+\left(\Omega_{j_{2}}^{0}\right)^{2}-\left(\Omega_{\ell}^{0}\right)^{2}\right)}{\Omega_{j_{2}}^{0}\Omega_{j_{1}}^{0}},\\
\end{align}to rewrite the radial function $J_{\ell j_{1}j_{2}\omega;\alpha_{1}\beta_{2}}^{\gamma}\left(r,r_{i},r_{\mathrm{src}}\right) $ as 
\begin{align}
J_{\ell j_{1}j_{2}\omega;\alpha_{1}\beta_{2}}^{\gamma}\left(r,r_{i},r_{\mathrm{src}}\right) & =-2i\omega\rho N_{\ell}^{j_{1}j_{2}}C_{j_{1}0j_{2}-\gamma}^{\ell-\gamma}\,\left(\frac{\Omega_{j_{2}}^{0}}{r}\right)^{\left|\gamma\right|}\mathcal{G}_{\ell j_{1}j_{2}\omega;\alpha_{1}\beta_{2}}^{\gamma}\left(r,r_{i},r_{\mathrm{src}}\right),\label{eq:G_radial}
\end{align}
and list the values of $\mathcal{G}_{\ell j_{1}j_{2};\alpha_{1}\beta_{2}}^{0}\left(r,r_{i},r_{\mathrm{src}}\right)$ and $\mathcal{G}_{\ell j_{1}j_{2};\alpha_{1}\beta_{2}}^{1}\left(r,r_{i},r_{\mathrm{src}}\right)$ in Table \ref{tab:Expressions-for-J}. We note that $\mathcal{G}_{\ell j_{1}j_{2};\alpha_{1}\beta_{2}}^{-1}=\mathcal{G}_{\ell j_{1}j_{2};\alpha_{1}\beta_{2}}^{1}$. We list the Clebsch-Gordan relations involved in the evaluation of the terms $\mathcal{G}_{\ell j_{1}j_{2}\omega;\alpha_{1}\beta_{2}}^{\gamma}$ in Appendix \ref{sec:Appendix_VSH-triple-integral}.

\begin{table}
\renewcommand{\arraystretch}{1.5}%
\begin{tabular}{|c|c|}
\hline 
Term & Expression\tabularnewline
\hline 
\hline 
$\mathcal{G}_{\ell j_{1}j_{2}\omega;\alpha_{1}\beta_{2}}^{0}\left(r,r_{i},r_{\mathrm{src}}\right)$ & $\gfnomega 0{\alpha_{1}}{j_{1}}\left(r,r_{i}\right)\frac{d}{dr}\gfnomega 0{\beta_{2}}{j_{2}}\left(r,r_{\mathrm{src}}\right)+\zeta_{\ell}^{j_{1}j_{2}}\gfnomega 1{\alpha_{1}}{j_{1}}\left(r,r_{i}\right)\frac{d}{dr}\gfnomega 1{\beta_{2}}{j_{2}}\left(r,r_{\mathrm{src}}\right)$\tabularnewline
\hline 
$\mathcal{G}_{\ell j_{1}j_{2}\omega;\alpha_{1}\beta_{2}}^{1}\left(r,r_{i},r_{\mathrm{src}}\right)$ & $\begin{array}{c}
\gfnomega 0{\alpha_{1}}{j_{1}}\left(r,r_{i}\right)\gfnomega 0{\beta_{2}}{j_{2}}\left(r,r_{\mathrm{src}}\right)+\zeta_{\ell}^{j_{1}j_{2}}\gfnomega 1{\alpha_{1}}{j_{1}}\left(r,r_{i}\right)\gfnomega 1{\beta_{2}}{j_{2}}\left(r,r_{\mathrm{src}}\right)\\
-\frac{1}{\Omega_{j_{1}}^{0}}\gfnomega 1{\alpha_{1}}{j_{1}}\left(r,r_{i}\right)\gfnomega 0{\beta_{2}}{j_{2}}\left(r,r_{\mathrm{src}}\right)-\frac{1}{\Omega_{j_{2}}^{0}}\gfnomega 0{\alpha_{1}}{j_{1}}\left(r,r_{i}\right)\gfnomega 1{\beta_{2}}{j_{2}}\left(r,r_{\mathrm{src}}\right)
\end{array}$\tabularnewline
\hline 
\end{tabular}

\caption{Expressions for $G_{\ell j_{0}j_{2};\alpha_{1}\beta_{2}}^{\gamma}\left(r,r_{i},r_{\mathrm{src}}\right)$ that contribute towards the change in the Green function in the PB VSH basis.\label{tab:Expressions-for-J}}

\renewcommand{\arraystretch}{1}
\end{table}

\subsection{Change in the cross-covariance}

The presence of flows in the background model alters properties of seismic waves such as the local propagation speed. Such a difference manifests itself in the surface measurements of wave velocity, and consequently in the two-point cross-covariances. We may express the resultant change in the cross-covariance in terms of the changes in the Green function as
\begin{equation}
\delta C\left(\mathbf{x}_{1},\mathbf{x}_{2};\omega\right)=\bm{l}\left(\mathbf{x}_{1}\right)\bm{l}\left(\mathbf{x}_{2}\right):\delta\mathbf{C}\left(\mathbf{x}_{1},\mathbf{x}_{2};\omega\right),\label{eq:dC}
\end{equation}where
\begin{align}
\delta\mathbf{C}\left(\mathbf{x}_{1},\mathbf{x}_{2};\omega\right) & =\omega^{2}P\left(\omega\right)\int d\Omega_{\mathrm{src}}\,\left[\delta\mathbf{G}_{r}^{*}\left(\mathbf{x}_{1},\mathbf{x}_{\mathrm{src}};\omega\right)\mathbf{G}_{r}\left(\mathbf{x}_{2},\mathbf{x}_{\mathrm{src}};\omega\right)+\left(1\leftrightarrow2\right)^{\dagger}\right],\label{eq:deltaC}
\end{align}and the subscript $r$ indicates that the second index of $\mathbf{G}$ is chosen to coincide with the radial direction at $\mathbf{x}_{\mathrm{src}}$. The term $\left(1\leftrightarrow2\right)^{\dagger}$ is obtained by switching the observation points $\mathbf{x}_{1}$ and $\mathbf{x}_{2}$ in the first term, followed by evaluating its conjugate-transpose. Substituting Equation (\ref{eq:dG_PB_VSH}) into Equation (\ref{eq:deltaC}) and integrating over the angular distribution of the sources, we obtain
\begin{equation}
\delta\mathbf{C}\left(\mathbf{x}_{1},\mathbf{x}_{2};\omega\right)=\sum_{\ell m\gamma}\int r^{2}dr\,u_{\ell m}^{\gamma}\left(r\right)\,\sum_{j_{1}j_{2}}\sum_{\alpha_{1}\alpha_{2}}\mathcal{C}_{\ell j_{1}j_{2}\omega;\alpha_{1}\alpha_{2}}^{\gamma}\left(r,r_{1},r_{2},r_{\mathrm{src}}\right)\PB{j_{1}j_{2},\alpha_{1}\alpha_{2}}{\ell m}\left(\hat{n}_{1},\hat{n}_{2}\right),\label{eq:dC_biposh}
\end{equation}where 
\begin{align}
\mathcal{C}_{\ell j_{1}j_{2}\omega;\alpha_{1}\alpha_{2}}^{\gamma}\left(r,r_{1},r_{2},r_{\mathrm{src}}\right) & =\omega^{2}P\left(\omega\right)\left(J_{\ell j_{1}j_{2}\omega;\alpha_{1}0}^{-\gamma*}\left(r,r_{1},r_{\mathrm{src}}\right)\gfnomega{\alpha_{2}}0{j_{2}}\left(r_{2},r_{\mathrm{src}}\right)\right.\nonumber \\
 & \left.\quad+\gfnomega{\alpha_{1}*}0{j_{1}}\left(r_{1},r_{\mathrm{src}}\right)J_{\ell j_{2}j_{1}\omega;\alpha_{2}0}^{\gamma}\left(r,r_{2},r_{\mathrm{src}}\right)\right),\label{eq:C_components_defn}
\end{align}
and $J_{\ell j_{1}j_{2}\omega;\alpha0}^{\gamma}$ is defined in Equation (\ref{eq:Jgamma_defn}). We derive the expression in Equation (\ref{eq:deltaC}) in Appendix \ref{sec:Appendix_source_angle_int}. The function $\mathcal{C}_{\ell j_{1}j_{2}\omega;\alpha_{1}\alpha_{2}}^{\gamma}$ obeys the symmetry relations in Equation \eqref{eq:Jgamma_symemtry}, as well as
\begin{equation}
\mathcal{C}_{\ell j_{1}j_{2}\omega;\alpha_{1}\alpha_{2}}^{\gamma}\left(r,r_{1},r_{2},r_{\mathrm{src}}\right)=\mathcal{C}_{\ell j_{2}j_{1}\omega;\alpha_{2}\alpha_{1}}^{-\gamma*}\left(r,r_{2},r_{1},r_{\mathrm{src}}\right).
\end{equation}
We define the line-of-sight-projected bipolar spherical harmonic 
\begin{equation}
P_{\ell m}^{j_{1}j_{2},\alpha_{1}\alpha_{2}}\left(\mathbf{x}_{1},\mathbf{x}_{2}\right)=\bm{l}\left(\mathbf{x}_{1}\right)\bm{l}\left(\mathbf{x}_{2}\right):\PB{j_{1}j_{2},\alpha_{1}\alpha_{2}}{\ell m}\left(\hat{n}_{1},\hat{n}_{2}\right),
\end{equation}
and collect the terms summed over in Equation (\ref{eq:dC_biposh}) to define
\begin{equation}
\mathcal{C}_{\ell m}^{\gamma}\left(r,\mathbf{x}_{1},\mathbf{x}_{2};\omega\right)=\sum_{j_{1}j_{2}}\sum_{\alpha_{1}\alpha_{2}}\mathcal{C}_{\ell j_{1}j_{2}\omega;\alpha_{1}\alpha_{2}}^{\gamma}\left(r,r_{1},r_{2},r_{\mathrm{src}}\right)P_{\ell m}^{j_{1}j_{2},\alpha_{1}\alpha_{2}}\left(\mathbf{x}_{1},\mathbf{x}_{2}\right)
\end{equation}
in order to simplify the notation. We rewrite Equation \eqref{eq:dC} in terms of this as 
\begin{equation}
\delta C\left(\mathbf{x}_{1},\mathbf{x}_{2};\omega\right)=\sum_{\ell m\gamma}\int r^{2}dr\,u_{\ell m}^{\gamma}\left(r\right)\,\mathcal{C}_{\ell m}^{\gamma}\left(r,\mathbf{x}_{1},\mathbf{x}_{2};\omega\right).\label{eq:dC_integrated}
\end{equation}

\section{Sensitivity kernel\label{subsec:Kernels}}

A change in the cross-covariance of seismic waves by $\delta C\left(\mathbf{x}_{1},\mathbf{x}_{2},\omega\right)$ as measured at the points $\mathbf{x}_1$ and $\mathbf{x}_2$ in turn results in a variation in the time $\tau_{12}$ that the wave takes to travel between these two points. At a linear order, this change in travel-time $\delta\tau_{12}$ may be related to the change in the cross-covariance through
\begin{equation}
\delta\tau_{12}=\int_{0}^{\infty}\frac{d\omega}{2\pi}\,2\Re\left[h^{*}\left(\mathbf{x}_{1},\mathbf{x}_{2},\omega\right)\delta C\left(\mathbf{x}_{1},\mathbf{x}_{2},\omega\right)\right],
\label{eq:dtau_freqdomain}
\end{equation}
 \citep{2002ApJ...571..966G}. Substituting Equation (\ref{eq:dC_integrated}) into Equation (\ref{eq:dtau_freqdomain}), we obtain a relation between the travel-time shifts and the components of the background flow velocity field:
\begin{equation}
\delta\tau_{12}=\sum_{\ell m\gamma}\int r^{2}dr\,K_{\gamma,\ell m}\left(r,\mathbf{x}_{1},\mathbf{x}_{2}\right)u_{\ell m}^{\gamma}\left(r\right),\label{eq:travel-time-kernel_velocity-integral}
\end{equation}
where $K_{\gamma,\ell m}\left(r;\mathbf{x}_{1},\mathbf{x}_{2}\right)$, defined as
\begin{equation}
K_{\gamma,\ell m}\left(r;\mathbf{x}_{1},\mathbf{x}_{2}\right)=\int_{0}^{\infty}\frac{d\omega}{2\pi}\,\left[h^{*}\left(\mathbf{x}_{1},\mathbf{x}_{2},\omega\right)\mathcal{C}_{\ell m}^{\gamma}\left(r,\mathbf{x}_{1},\mathbf{x}_{2};\omega\right)+h\left(\mathbf{x}_{1},\mathbf{x}_{2},\omega\right)\left(-1\right)^{m}\mathcal{C}_{\ell-m}^{-\gamma*}\left(r,\mathbf{x}_{1},\mathbf{x}_{2};\omega\right)\right],\label{eq:kernel_components}
\end{equation}
is the covariant component of the sensitivity kernels corresponding to the component of the flow velocity denoted by $\gamma$ in the PB VSH basis. We see that $K_{-\gamma,\ell-m}=\left(-1\right)^{m}K_{\gamma,\ell m}^{*}$, reaffirming the vector nature of the kernel. Specifically, we find that the components of the kernel for $\gamma=-1$ and $\gamma=1$ are related through $K_{-1,\ell-m}=\left(-1\right)^{m}K_{1,\ell m}^{*}$. This also tells us that the kernel $K_{0,\ell0}$ --- which corresponds to axisymmetric radial flows --- is purely real. Equation \eqref{eq:travel-time-kernel_velocity-integral} sets up the inverse problem that we need to solve to compute the velocity components. We may further use the condition $u_{\ell m}^{\gamma*} = (-1)^m u_{\ell-m}^{-\gamma}$ --- arising from the fact that the velocity $\mathbf{u}(\mathbf{x})$ is real --- to limit the number of terms that appear in Equation (\ref{eq:travel-time-kernel_velocity-integral}).

We may use the symmetry relations from Equation (\ref{eq:Jgamma_symemtry}) to obtain the expression for $K_{\gamma,\ell m}\left(r;\mathbf{x}_{1},\mathbf{x}_{2}\right)$ in the PB VSH basis to be 
\begin{align}
K_{\gamma,\ell m}\left(r;\mathbf{x}_{1},\mathbf{x}_{2}\right) & =\int_{0}^{\infty}\frac{d\omega}{2\pi}\,\sum_{j_{1}j_{2}}\sum_{\alpha_{1}\alpha_{2}}\mathcal{K}_{\ell j_{1}j_{2}\omega;\alpha_{1}\alpha_{2}}^{\gamma}\left(r,\mathbf{x}_{1},\mathbf{x}_{2}\right)P_{\ell m}^{j_{1}j_{2},\alpha_{1}\alpha_{2}}\left(\mathbf{x}_{1},\mathbf{x}_{2}\right),\label{eq:K_sepvar}
\end{align}
where we have defined
\begin{align}
\mathcal{K}_{\ell j_{1}j_{2}\omega;\alpha_{1}\alpha_{2}}^{\gamma}\left(r,\mathbf{x}_{1},\mathbf{x}_{2}\right) & =2\Re\left[h^{*}\left(\mathbf{x}_{1},\mathbf{x}_{2},\omega\right)\mathcal{C}_{\ell j_{1}j_{2}\omega;\alpha_{1}\alpha_{2}}^{\gamma}\left(r,r_{1},r_{2};\omega\right)\right].\label{eq:K_components_defn}
\end{align}
The function $\mathcal{K}_{\ell j_{1}j_{2}\omega;\alpha_{1}\alpha_{2}}^{\gamma}$ satisfies symmetry relations analogous to $\mathcal{C}_{\ell j_{1}j_{2}\omega;\alpha_{1}\alpha_{2}}^{\gamma}$. Specifically, we use
\begin{equation}
\mathcal{K}_{\ell j_{1}j_{2}\omega;\alpha_{1}\alpha_{2}}^{-\gamma}=\left(-1\right)^{\ell+j_{1}+j_{2}}\mathcal{K}_{\ell j_{1}j_{2}\omega;\alpha_{1}\alpha_{2}}^{\gamma},
\end{equation}
to see that $\mathcal{K}_{\ell j_{1}j_{2}\omega;\alpha_{1}\alpha_{2}}^{0}$ is non-zero only for even values of $\ell+j_{1}+j_{2}$, and the combinations $\mathcal{K}_{\ell j_{1}j_{2}\omega;\alpha_{1}\alpha_{2}}^{1}+\mathcal{K}_{\ell j_{1}j_{2}\omega;\alpha_{1}\alpha_{2}}^{-1}$ and $\mathcal{K}_{\ell j_{1}j_{2}\omega;\alpha_{1}\alpha_{2}}^{1}-\mathcal{K}_{\ell j_{1}j_{2}\omega;\alpha_{1}\alpha_{2}}^{-1}$ are non-zero for even and odd values of $\ell+j_{1}+j_{2}$ respectively.

\begin{figure*}
\includegraphics[scale=0.85]{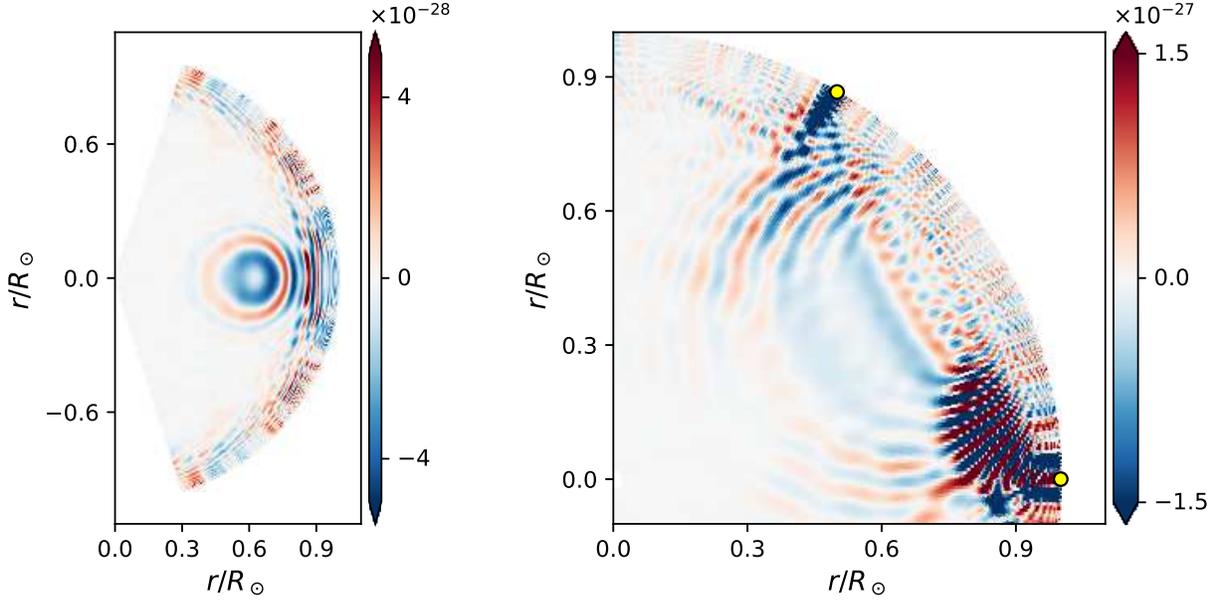}
\caption{\label{fig:Ku_3D} Longitudinal (left) and Equatorial (right) sections of the three-dimensional profile of the kernel for two observation points located on the Equator separated by 60 degrees at a height of $200$ km above the photosphere. The points have been marked in yellow in the panel on the right. The kernel has been multiplied by the radial sound-speed profile, and the color scale has been saturated to highlight the deeper layers. The kernel has been computed in CGS units.}
\end{figure*}

We may compute the three-dimensional profile of the kernel by summing up over the kernel components and using $K^\gamma_{\ell m}=K^*_{\gamma,\ell m}$ to obtain 
\begin{equation}
\mathbf{K}\left(\mathbf{x};\mathbf{x}_{1},\mathbf{x}_{2}\right)=\sum_{\gamma\ell m}K_{\gamma,\ell m}^{*}\left(r;\mathbf{x}_{1},\mathbf{x}_{2}\right)\PBlm{\gamma}\left(\hat{n}\right).\label{eq:K_3D}
\end{equation}We may use the expansion of the PB VSH in the spherical polar basis and obtain the appropriately directed components of the kernel to be
\begin{align}
K_{r}\left(\mathbf{x};\mathbf{x}_{1},\mathbf{x}_{2}\right) & =\sum_{\ell m}K_{0,\ell m}^{*}\left(r;\mathbf{x}_{1},\mathbf{x}_{2}\right)Y_{\ell m}\left(\hat{n}\right),\\
K_{\theta}\left(\mathbf{x};\mathbf{x}_{1},\mathbf{x}_{2}\right) & =-\sqrt{2}\sum_{\ell m}\Re\left[K_{1,\ell m}^{*}\left(r;\mathbf{x}_{1},\mathbf{x}_{2}\right)\sgshlm{+1}\left(\hat{n}\right)\right],\\
K_{\phi}\left(\mathbf{x};\mathbf{x}_{1},\mathbf{x}_{2}\right) & =\sqrt{2}\sum_{\ell m}\Im\left[K_{1,\ell m}^{*}\left(r;\mathbf{x}_{1},\mathbf{x}_{2}\right)\sgshlm{+1}\left(\hat{n}\right)\right].
\end{align}
We plot the cross-sections of the three-dimensional profile of the kernel in Figure \ref{fig:Ku_3D} choosing the observation points to be $\mathbf{x}_1=(R_\odot+200\,\mathrm{km},\pi/2,0)$ and $\mathbf{x}_2=(R_\odot+200\,\mathrm{km},\pi/2,\pi/3)$. The panel on the left shows a longitudinal slice through $\phi=\pi/6$ --- midway between the azimuths at which the measurements are carried out --- whereas the one on the right shows a latitudinal section through the Equator, passing through the observation points. The kernels have been computed by summing up over VSH modes of the flow velocity with angular degrees in the range $0\leq\ell\leq160$, where the upper bound arises from the limits of the numerical accuracy in evaluating Clebsch-Gordan coefficients. We describe the numerical evaluation in section \ref{sec:Numerical}.

The kernels for $m=0$ are of particular interest as these correspond to azimuthally symmetric flow profiles such as meridional flows and differential rotation. We shall look at these in the following sections.

\subsection{Kernels for axisymmetric flows}

Components $u_{\ell0}^{\left(\alpha\right)}\left(r\right)$ of axisymmetric flows in the Hansen VSH basis have a geometrical interpretation arising from the fact that the Hansen basis vector $\Hansen{\left(1\right)}{\ell0}\left(\hat{n}\right)$ is directed along $\mathbf{e}_{\theta}$ whereas $\Hansen{\left(0\right)}{\ell0}\left(\hat{n}\right)$ is directed along $\mathbf{e}_{\phi}$. This implies that spheroidal velocity profiles such as meridional flows may be expressed in terms of the two sets of components $u_{\ell0}^{\left(-1\right)}\left(r\right)$ and $u_{\ell0}^{\left(1\right)}\left(r\right)$ whereas toroidal profiles may be expressed in terms of $u_{\ell0}^{\left(0\right)}\left(r\right)$. We may use the relationship between the Hansen and the PB VSH bases from Equation (\ref{eq:PB_Hansen_conversion}) alongside the conjugation relation $u_{\ell0}^{\alpha*}=u_{\ell0}^{-\alpha}$ to obtain 
\begin{equation}
\begin{aligned}u_{\ell0}^{\left(1\right)}\left(r\right) & =\sqrt{2}\Re\left[u_{\ell0}^{1}\left(r\right)\right],\\
u_{\ell0}^{\left(0\right)}\left(r\right) & =-\sqrt{2}i\Im\left[u_{\ell0}^{1}\left(r\right)\right].
\end{aligned}
\label{eq:u_axisym_comp_PB_Hansen}
\end{equation}
This further implies that the tangential components of axisymmetric flows may be expanded in terms of just the PB VSH components $u_{\ell0}^{1}\left(r\right)$. We develop the following analysis in terms of the real and imaginary components of $u_{\ell0}^{1}\left(r\right)$ to demonstrate that the kernels are manifestly real.

We may use $K_{-1,\ell0}=K_{1,\ell0}^{*}$ and rewrite the expression for the travel-time shift from Equation (\ref{eq:travel-time-kernel_velocity-integral})
in the form 
\begin{equation}
\delta\tau_{12}=\sum_{\ell}\int r^{2}dr\,\left[K_{0,\ell0}\left(r\right)u_{\ell0}^{0}\left(r\right)+2\Re\left[K_{1,\ell0}\left(r\right)\right]\Re\left[u_{\ell0}^{1}\left(r\right)\right]-2\Im\left[K_{1,\ell0}\left(r\right)\right]\Im\left[u_{\ell0}^{1}\left(r\right)\right]\right],\label{eq:dtau_axisym_PB}
\end{equation}where we have suppressed the explicit dependence of the kernel components on the observation points $\mathbf{x}_{1}$ and $\mathbf{x}_{2}$ for brevity. We define
\begin{equation}
\begin{aligned}K_{r,\ell0}\left(r\right) & =K_{0,\ell0}\left(r\right),\\
K_{\theta,\ell0}\left(r\right) & =2\Re\left[K_{1,\ell0}\left(r\right)\right],\\
K_{\phi,\ell0}\left(r\right) & =-2\Im\left[K_{1,\ell0}\left(r\right)\right],
\end{aligned}
\label{eq:K_r_theta_phi_l0}
\end{equation}
and rewrite Equation (\ref{eq:dtau_axisym_PB}) as \begin{equation}
\delta\tau_{12}=\sum_{\ell}\int_{0}^{R_{\odot}}r^{2}dr\,\left[K_{r,\ell0}\left(r\right)u_{\ell0}^{0}\left(r\right)+K_{\theta,\ell0}\left(r\right)\Re\left[u_{\ell0}^{1}\left(r\right)\right]+K_{\phi,\ell0}\left(r\right)\Im\left[u_{\ell0}^{1}\left(r\right)\right]\right].\label{eq:dtau_axisym_PB_geom}
\end{equation}
The first terms in the expression corresponds to a radial flow, the second to a poloidal flow, whereas the last term corresponds to a toroidal flow. We may use information about the geometrical orientations of the flow field --- if available --- to further restrict the number of coefficients. We also note that the components of the kernel as defined here are related to those in the Hansen basis through a scaling.

\subsection{Kernels for meridional flows}

Meridional flows are restricted to the $\mathbf{e}_{r}-\mathbf{e}_{\theta}$ plane by definition, and are assumed to be azimuthally symmetric. Under these assumptions we need to solve only for the $m=0$ component, and may further use the fact that the flow components are real and satisfy $u_{\ell0}^{+1}=u_{\ell0}^{-1}$. Equation (\ref{eq:dtau_axisym_PB_geom}) tells us that a change in travel time may be related to the flow coefficients through 
\begin{align}
\delta\tau_{12} & =\sum_{\ell}\int r^{2}dr\,\left[K_{r,\ell0}\left(r\right)u_{\ell0}^{0}\left(r\right)+K_{\theta,\ell0}\left(r\right)u_{\ell0}^{1}\left(r\right)\right].
\end{align}
We therefore need to compute the components $K_{r,\ell0}\left(r;\mathbf{x}_{1},\mathbf{x}_{2}\right)$ and $K_{\theta,\ell0}\left(r;\mathbf{x}_{1},\mathbf{x}_{2}\right)$. We use Equations (\ref{eq:K_sepvar}) and (\ref{eq:K_r_theta_phi_l0}) to obtain 
\begin{align}
K_{r,\ell0}\left(r;\mathbf{x}_{1},\mathbf{x}_{2}\right) & =\int_{0}^{\infty}\frac{d\omega}{2\pi}\,\sum_{j_{1}j_{2}}\sum_{\alpha_{1}\alpha_{2}}\mathcal{K}_{\ell j_{1}j_{2}\omega;\alpha_{1}\alpha_{2}}^{0}\left(r,\mathbf{x}_{1},\mathbf{x}_{2}\right)\Re\left[P_{\ell0}^{j_{1}j_{2},\alpha_{1}\alpha_{2}}\left(\mathbf{x}_{1},\mathbf{x}_{2}\right)\right],\label{eq:Kr_l0}\\
K_{\theta,\ell0}\left(r;\mathbf{x}_{1},\mathbf{x}_{2}\right) & =\int_{0}^{\infty}\frac{d\omega}{2\pi}\,\sum_{j_{1}j_{2}}\sum_{\alpha_{1}\alpha_{2}}\left(1+\left(-1\right)^{\ell+j_{1}+j_{2}}\right)\mathcal{K}_{\ell j_{1}j_{2}\omega;\alpha_{1}\alpha_{2}}^{1}\left(r,\mathbf{x}_{1},\mathbf{x}_{2}\right)\Re\left[P_{\ell0}^{j_{1}j_{2},\alpha_{1}\alpha_{2}}\left(\mathbf{x}_{1},\mathbf{x}_{2}\right)\right],\label{eq:Ktheta_l0}
\end{align}
We find that the contributions towards $K_{1,\ell0}^{\theta}$ comes only from the modes for which $\ell+j_{1}+j_{2}$ is even. The same constraint also implicitly holds for $K_{r,\ell0}$ as $\mathcal{K}_{\ell j_{1}j_{2}\omega;\alpha_{1}\alpha_{2}}^{0}$ is non-zero only for even values of $\ell+j_{1}+j_{2}$. The geometric orientation of the flow field would further reduce the number of $\ell$s contributing towards the travel time, for example meridional flows may be represented in terms of even $\ell$s.

\begin{figure}
\includegraphics[scale=1]{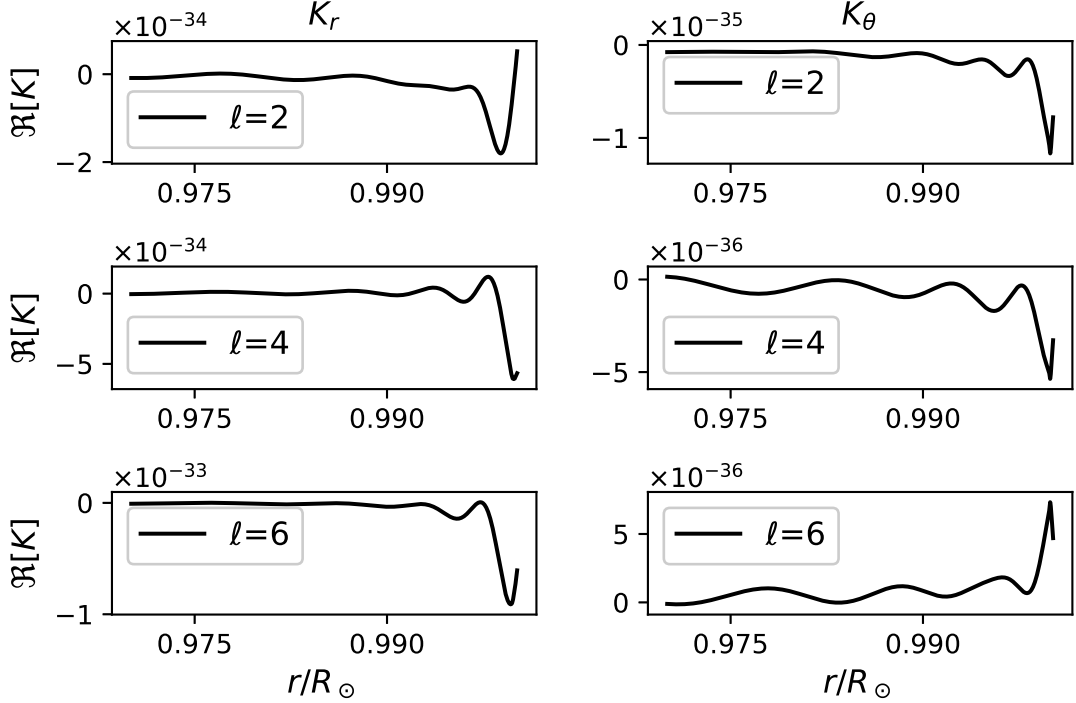}
\caption{Kernels for radial and tangential components of meridional flow for two observation points at $\mathbf{x}_1=(R_\odot+200\,\mathrm{km},\pi/2,0)$ and $\mathbf{x}_2=(R_\odot+200\,\mathrm{km},\pi/4,0)$. The kernels are in units of $\mathrm{s}/(\mathrm{cm}/\mathrm{s})/\mathrm{cm}^3$.}
\label{fig:Kernels-merid_flow}
\end{figure}

We plot $K_{r,\ell0}$ and $K_{\theta,\ell0}$ for different values of $\ell$ in Figure \ref{fig:Kernels-merid_flow}, choosing the observation points to be $\mathbf{x}_1=(R_\odot+200\,\mathrm{km},\pi/2,0)$ and $\mathbf{x}_2=(R_\odot+200\,\mathrm{km},\pi/4,0)$. We may simplify the inverse problem further if we assume mass conservation, and solve for kernels corresponding to the $\phi$-component of the stream function. We describe this procedure in Section \ref{subsec:Mass-conservation:-kernels}.

\subsection{Kernels for rotation \label{subsec:Kernels-for-rotation}}

\begin{figure*}
\includegraphics{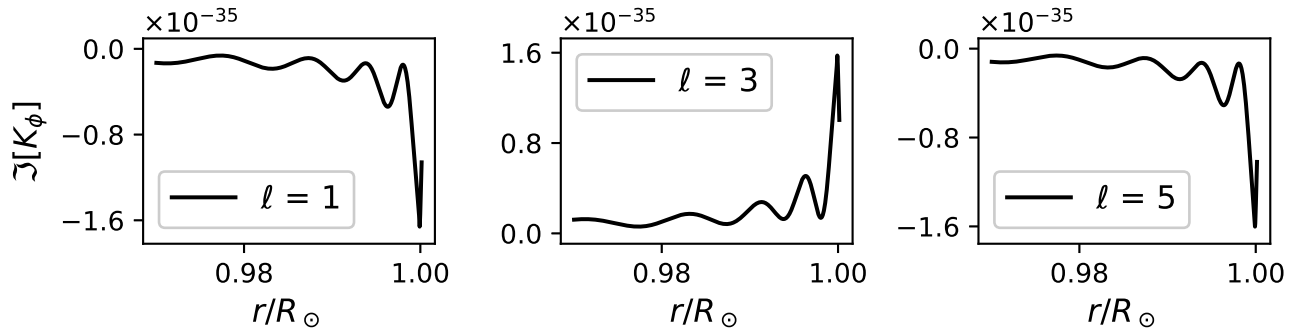}
\caption{Radial profiles of the imaginary part of the kernels for rotation for two observation points at $\mathbf{x}_1=(R_\odot+200\,\mathrm{km},\pi/2,0)$ and $\mathbf{x}_2=(R_\odot+200\,\mathrm{km},\pi/2,\pi/3)$, for various spherical harmonic degrees $\ell$ and the azimuthal order $m=0$. The kernels are in units of $\mathrm{s}/(\mathrm{cm}/\mathrm{s})/\mathrm{cm}^3$.}
\label{fig:Kernels-for-rotation}
\end{figure*}

Rotations of the Sun may be assumed to azimuthally symmetric $\left(m=0\right)$ and directed along $\mathbf{e}_{\phi}$. In this case the flow components are imaginary and satisfy $u_{\ell0}^{+1}=-u_{\ell0}^{-1}$. This implies that we need to solve for the kernel functions $K_{\phi,\ell0}\left(r;\mathbf{x}_{1},\mathbf{x}_{2}\right)$ that relate a change in travel time to the background flow through
\begin{align}
\delta\tau_{12} & =\sum_{\ell}\int r^{2}dr\,K_{\phi,\ell0}\left(r;\mathbf{x}_{1},\mathbf{x}_{2}\right)\Im\left[u_{\ell0}^{1}\left(r\right)\right].
\end{align}
We may rewrite the expression for $K_{\phi,\ell0}\left(r;\mathbf{x}_{1},\mathbf{x}_{2}\right)$ as 
\begin{align}
K_{\phi,\ell0}\left(r;\mathbf{x}_{1},\mathbf{x}_{2}\right) & =\int_{0}^{\infty}\frac{d\omega}{2\pi}\,\sum_{j_{1}j_{2}}\sum_{\alpha_{1}\alpha_{2}}\left(\left(-1\right)^{\ell+j_{1}+j_{2}}-1\right)\mathcal{K}_{\ell j_{1}j_{2}\omega;\alpha_{1}\alpha_{2}}^{1}\left(r,\mathbf{x}_{1},\mathbf{x}_{2}\right)\Im\left[P_{\ell0}^{j_{1}j_{2},\alpha_{1}\alpha_{2}}\left(\mathbf{x}_{1},\mathbf{x}_{2}\right)\right].\label{eq:Kphi_l0}
\end{align}
We find that the contributions only come from the modes for which $\ell+j_{1}+j_{2}$ is odd. The transformation of the Hansen VSH under coordinate inversion indicates that we only need to solve for the coefficients $u_{\ell0}^{1}\left(r\right)$ for odd values of $\ell$ \citep{1991ApJ...369..557R}, with $\ell=1$ corresponding to uniform or radially differential rotation, and $\ell\geq3$ corresponding to latitudinal differential rotation. We compute the function $K_{\phi,\ell0}$ for the observation points $\mathbf{x}_1=(R_\odot+200\,\mathrm{km},\pi/2,0)$ and $\mathbf{x}_2=(R_\odot+200\,\mathrm{km},\pi/2,\pi/3)$, and plot their radial profiles in Figure \ref{fig:Kernels-for-rotation} for different values of $\ell$.

\subsection{Mass conservation: kernels for the stream function \label{subsec:Mass-conservation:-kernels}}

A temporally-stationary, mass-conserving flow field $\mathbf{u}\left(\mathbf{x}\right)$ satisfies the continuity relation $\grad\cdot\left(\rho\mathbf{u}\right)=0$, and may be represented in terms of a stream function $\bm{\psi}\left(\mathbf{x}\right)$ as 
\begin{equation}
\mathbf{u}\left(\mathbf{x}\right)=\frac{1}{\rho}\grad\times\bm{\psi}\left(\mathbf{x}\right).\label{eq:u_psi}
\end{equation}
The choice of stream function is not unique for a specified flow field $\mathbf{u}(\mathbf{x})$, as the transformation $\bm{\psi}\rightarrow\bm{\psi}+\grad\xi$ for a scalar field $\xi(\mathbf{x})$ leads the same flow velocity. This ambiguity may be eliminated by imposing a suitable constraint on $\bm{\psi}$, also referred to as gauge fixing. However this is not critical to our analysis, firstly because we are interested in the existence and not in the uniqueness of the stream function, and secondly because in the interesting special case of meridional flows, the stream function is toroidal, and consequently free from such an ambiguity.

We may evaluate the kernel for the stream function by substituting Equation \eqref{eq:u_psi} into 
$
\delta\tau=\int d\mathbf{x}\,\mathbf{K}_{\mathbf{u}}\left(\mathbf{x}\right)\cdot\mathbf{u}\left(\mathbf{x}\right)
$
and integrating by parts, to obtain

\begin{equation}
\delta\tau=\int d\mathbf{x}\left(\bm{\nabla}\times\left(\frac{1}{\rho}\mathbf{K}_{\mathbf{u}}\left(\mathbf{x}\right)\right)\right)\cdot\bm{\psi}\left(\mathbf{x}\right)-\int dS\left.\frac{1}{\rho}\mathbf{K}_{\mathbf{u}}\left(\mathbf{x}\right)\cdot\bm{\psi\left(\mathbf{x}\right)}\right\rfloor _{S},
\end{equation}
where we have suppressed the explicit dependence of the kernel on the observation points to simplify the notation. The second term is a surface integral over the boundary of the domain, and may be dropped if the stream function $\psi(\mathbf{x})$ goes to zero at the extremities. In such a case the kernel for the stream function is related to that for the flow through 
\begin{equation}
\mathbf{K}_{\psi}\left(\mathbf{x}\right)=\grad\times\left(\frac{1}{\rho}\mathbf{K}_{\mathbf{u}}\left(\mathbf{x}\right)\right).
\label{Kpsi_Ku_vec}
\end{equation}
We may split Equation \eqref{Kpsi_Ku_vec} into components in the PB VSH basis as
\begin{align}
K_{\psi,0,\ell m}\left(r\right) & =-\frac{i\Omega_{\ell}^{0}}{\rho r}\left(K_{-1,\ell m}\left(r\right)-K_{+1,\ell m}\left(r\right)\right),\\
K_{\psi,\pm1,\ell m}\left(r\right) & =\pm i\left(\frac{1}{r}\frac{d}{dr}\left(\frac{r K_{\pm1,\ell m}\left(r\right)}{\rho}\right)-\frac{\Omega_{\ell}^{0}}{\rho r}K_{0,\ell m}\left(r\right)\right).
\end{align}
\citep[see][ for the components of the curl]{DahlenTromp}. In the special case of meridional flow --- where the velocity field is entirely in the $\mathbf{e}_{r}-\mathbf{e}_{\theta}$ plane --- the stream function is directed along $\mathbf{e}_{\phi}$. In addition, an axisymmetric flow field would necessitate a stream function that is azimuthally symmetric as well. Drawing an analogy with section \ref{subsec:Kernels-for-rotation} and using $\psi_{\ell0}^{+1}\left(r\right)=-\psi_{\ell0}^{-1}\left(r\right)$, we compute the kernel component 
\begin{align}
K_{\psi_{\phi},\ell0}\left(r\right) & = -\frac{1}{r}\frac{d}{dr}\left(\frac{r K_{\theta,\ell0}\left(r\right)}{\rho}\right) + 2\Omega_{\ell}^{0}\frac{K_{r,\ell0}\left(r\right)}{\rho r}.
\end{align}
A change in travel time would be related to the stream function component $\psi_{\ell0}^{+1}$ through 
\begin{align}
\delta\tau_{12} & =\sum_{\ell}\int_{0}^{R_{\odot}}r^{2}dr\,K_{\psi_{\phi},\ell0}\left(r;\mathbf{x}_{1},\mathbf{x}_{2}\right)\Im\left[\psi_{\ell0}^{+1}\left(r\right)\right].
\end{align}
Once we evaluate the stream function, we may compute the flow coefficients from it using 
\begin{align}
u_{\ell0}^{0}\left(r\right) & =\frac{2\Omega_{\ell}^{0}}{\rho r}\Im\left[\psi_{\ell0}^{+1}\left(r\right)\right],\quad u_{\ell0}^{\pm1}\left(r\right)=\frac{1}{\rho r}\frac{d}{dr}\left(r \Im\left[\psi_{\ell0}^{+1}\left(r\right)\right]\right).
\end{align}
We may further compute the flow velocity in spherical polar coordinates as

\begin{equation}
\begin{aligned}u_{r}\left(\mathbf{x}\right) & =\sum_{\ell}u_{\ell0}^{0}\left(r\right)\,Y_{\ell0}\left(\hat{n}\right),\\
u_{\theta}\left(\mathbf{x}\right) & =\sum_{\ell}\frac{1}{\Omega_{\ell}^{0}}\,u_{\ell0}^{+1}\left(r\right)\,\partial_{\theta}Y_{\ell0}\left(\hat{n}\right).
\end{aligned}
\end{equation}

We demonstrate that this approach reproduces the standard spherical-polar coordinate results by choosing the specific example of meridional flows, for which the stream function is axisymmetric and directed along $\mathbf{e}_\phi$. Such a flow is more conveniently analysed in the Hansen VSH basis. We note that for $m=0$, the Hansen basis vector $\Hansen{(0)}{\ell 0}(\hat{n}) = -i\mathbf{e}_\phi \partial_\theta Y_{\ell 0}(\hat{n})/\sqrt{(\ell(\ell + 1))}$. The azimuthal component of the stream function may therefore be represented as
\begin{equation}
\psi(\mathbf{x}) = \mathbf{e}_\phi\cdot\bm{\psi}(\mathbf{x})= \sum_{\ell}\Im\left[\psi^{(0)}_{\ell0}(r)\right] \frac{1}{\sqrt{\ell(\ell+1)}}\partial_{\theta}Y_{\ell0}(\hat{n}),
\label{eq:psi_legendre}
\end{equation}
where the components $\psi^{(0)}_{\ell 0}$ are related to the PB-basis components $\psi_{\ell 0}^{1}$ through $\psi^{(0)}_{\ell 0} = -\sqrt{2}\psi_{\ell 0}^{+1}$. To simplify the notation, we define $\psi_\ell(r)=\Im[\psi^{(0)}_{\ell 0}(r)]$ The flow velocity for meridional circulation may be expressed in the Hansen basis as 

\begin{equation}
\rho\mathbf{u}\left(\mathbf{x}\right)=-\sum_{\ell}\sqrt{2}\Omega_{\ell}^{0}\frac{\psi_{\ell}(r)}{r}Y_{\ell0}\left(\hat{n}\right)\mathbf{e}_{r}+\frac{1}{\sqrt{2}\Omega_{\ell}^{0} r}\frac{d\left(r\psi_{\ell}(r)\right)}{dr}\,\partial_{\theta}Y_{\ell0}\left(\hat{n}\right)\,\mathbf{e}_{\theta}.\label{eq:u_psi_meridional}
\end{equation}
On the other hand, Equation \eqref{eq:u_psi} may be expanded in spherical polar coordinates to

\begin{equation}
\mathbf{u}\left(\mathbf{x}\right)=\frac{1}{\rho r\sin\theta}\partial_{\theta}\left(\psi\left(\mathbf{x}\right)\sin\theta\right)\mathbf{e}_{r}-\frac{1}{\rho r}\partial_{r} \left(r \psi\left(\mathbf{x}\right)\right)\,\mathbf{e}_{\theta}.
\label{eq:u_psi_meridional_spherical}
\end{equation}

We may substitute Equation \eqref{eq:psi_legendre} into Equation \eqref{eq:u_psi_meridional_spherical}, and use the fact that $Y_{\ell 0}(\hat{n})$ are the eigenfunction of the Laplacian on a sphere corresponding to an eigenvalue of $-\ell(\ell+1)$, to reproduce Equation \eqref{eq:u_psi_meridional}. This demonstrates that an inversion for the stream function is equivalent to solving for the radial components $\psi^{(0)}_{\ell 0}$ (or equivalently $\psi^{+1}_{\ell 0}$). Such an approach had been used by \citet{2015ApJ...813..114R} and \citet{2018ApJ...863...39M} to invert for meridional circulation.

We further demonstrate that the travel times computed using the stream function are identical to that computed using the flow by choosing a specific model of the stream function. We retain only the term corresponding to $\ell=2$ in Equation \eqref{eq:psi_legendre}, and choose the radial function $\psi_2(r)$ to be of the form 

\begin{equation}
\psi_{2}\left(r\right)=A\,\rho\left(r\right)\exp\left(-\frac{\left(r-r_{0}\right)^{2}}{2\sigma^{2}}\right)d\left(r\right),
\label{eq:psi2}
\end{equation}
where $r_0 = 0.87R_\odot$, $\sigma = 0.05R_\odot$, the amplitude $A$ chosen to produce a maximum horizontal surface velocity of $20\,\mathrm{m}/\mathrm{s}$, and the function $d(r)$ being a decay term that ensures that the stream function falls to zero beyond the solar surface.

\begin{figure}
    \includegraphics{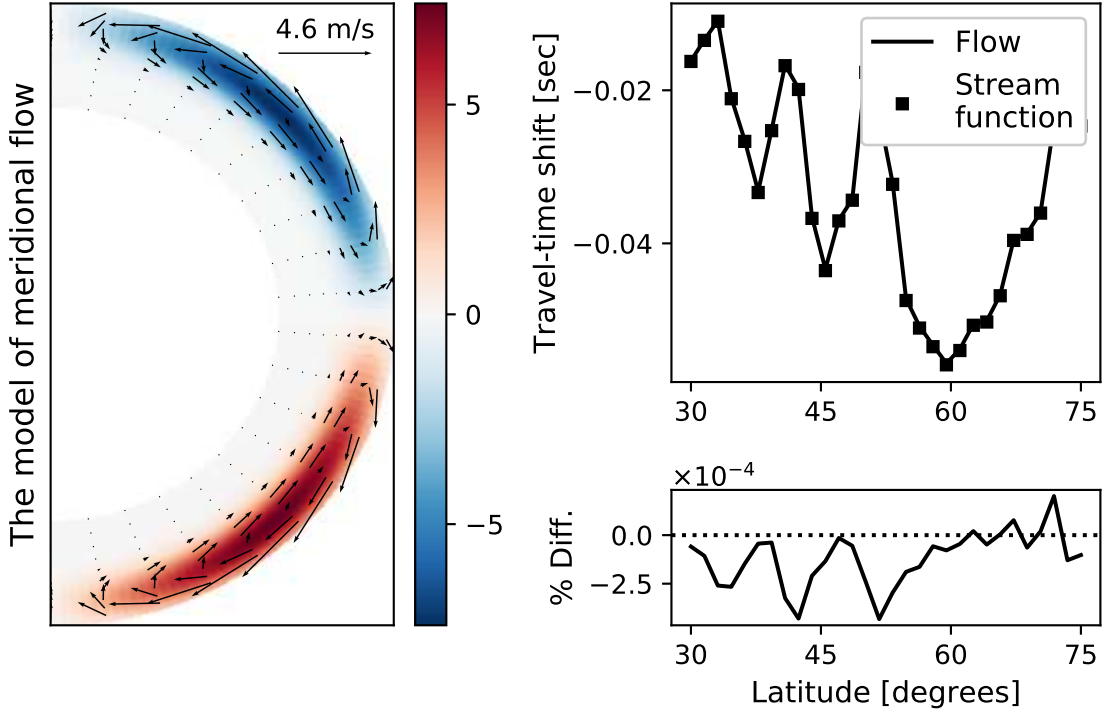}
    \caption{Left: Longitudinal cross-section of the stream function described in Equation \eqref{eq:psi2}. The color indicates the magnitude of the $\phi$-component of the stream function, and the arrows indicate the corresponding flow velocity. Right top: Travel-time shifts experienced by seismic waves traversing through the flow in the left panel, measured between two points on the same longitude, one at the Equator and the other located at various latitudes in the northern hemisphere. The line represents measurements using the kernel for flows, whereas the squares represent the same measurement but using the kernel for the stream function. Right bottom: Relative difference between the travel-time shifts in the top panel.}
    \label{fig:streamfn_traveltimes}
\end{figure}

We plot the travel time shifts obtained between the points $\mathbf{x}_1=(R_\odot+200\,\mathrm{km},\pi/2,0)$ and $\mathbf{x}_2=(R_\odot+200\,\mathrm{km},\theta,0)$ for several choices of the co-latitude $\theta$ in Figure \ref{fig:streamfn_traveltimes}. We find that there is a reasonable agreement between the travel-time shifts computed using the two approaches.

The number of parameters may be further reduced by representing the stream functions components $\psi_{\ell0}^{+1}\left(r\right)$ in a B-spline basis, for example as used by \citet{2018ApJ...863...39M}. This might lead to a significant simplification of inverse problems for meridional flows, as well as make them them better posed.

\subsection{Validating kernels for uniform rotation}

We verify our result for the kernel by comparing the wave travel times computed using two approaches: the first where we look at the change in cross-covariances arising in a rotating frame, and secondly where we treat the rotation as a flow about a steady background and evaluate the travel-time shift using Equation (\ref{eq:travel-time-kernel_velocity-integral}). We make the assumption that the cross-covariance is being measured between waves at two points on the equator separated azimuthally by $\Delta\phi$, both the points lying at the same observation radius $r_{\mathrm{obs}}$. We also leave out line-of-sight projections for algebraic simplicity. The cross-covariance in a frame rotating uniformly about the $\hat{z}$-axis at an angular speed $\Omega_{\mathrm{rot}}$ is related to that in a fixed frame through
\begin{equation}
C_{\mathrm{rotating}}\left(r_{\mathrm{obs}},\Delta\phi,t\right)=C_{\mathrm{fixed}}\left(r_{\mathrm{obs}},\Delta\phi-\Omega_{\mathrm{rot}}t,t\right).\label{eq:C_rot}
\end{equation}
The frequency-domain way of looking at the same would be a Doppler shift arising due to a uniformly-moving receiver. The difference in cross-covariances leads to a difference in measured travel times given by 
\begin{align}
\delta\tau\left(\Delta\phi\right) & =\int dt\,h\left(t\right)\left(C_{\mathrm{rotating}}\left(r_{\mathrm{obs}},\Delta\phi,t\right)-C_{\mathrm{fixed}}\left(r_{\mathrm{obs}},\Delta\phi,t\right)\right).\label{eq:dtau_cc}
\end{align}
On the other hand, treating the uniform solid-body rotation as a flow leads to a velocity field $\mathbf{u}\left(\mathbf{x}\right)=\Omega_{\mathrm{rot}}r\sin\theta\,\mathbf{e}_{\phi}$. We may express this in the PB VSH basis as 
\begin{equation}
\mathbf{u}\left(\mathbf{x}\right)=\sqrt{\frac{4\pi}{3}}i\Omega_{\mathrm{rot}}r\left(\PB{+1}{10}\left(\theta,\phi\right)-\PB{-1}{10}\left(\theta,\phi\right)\right),\label{eq:u_10_PB_VSH}
\end{equation}
We see the only non-zero spherical harmonic components correspond to $\ell=1$ and $m=0$. The shift in travel times in the first Born approximation may be obtained from Equation (\ref{eq:dtau_axisym_PB_geom}) as 
\begin{align}
\delta\tau_{12} & =\int_{0}^{R_{\odot}}r^{2}dr\,K_{\phi,10}\left(r;\mathbf{x}_{1},\mathbf{x}_{2}\right)\Im\left[u_{10}^{+1}\left(r\right)\right],\label{eq:dtau_uniform_kernel}
\end{align}
where the kernel $K_{\phi,10}\left(r;\mathbf{x}_{1},\mathbf{x}_{2}\right)$ is obtained by substituting $\ell=1$ in Equation (\ref{eq:Kphi_l0}). We find that that the only contribution to  $K_{\phi,10}\left(r;\mathbf{x}_{1},\mathbf{x}_{2}\right)$ comes from the modes corresponding to $j_{2}=j_{1}$, and we drop the subscript and use the symbol $j$ in subsequent analysis to refer
to the contributing wave modes. There is no contribution from $j=0$ as $j_{1}=j_{2}=0$ would restrict $\ell$ to $0$. The angular function $P_{10}^{jj,00}\left(\mathbf{x}_{1},\mathbf{x}_{2}\right)$ is equal
to the bipolar spherical harmonic $Y_{10}^{jj}\left(\hat{n}_{1},\hat{n}_{2}\right)$, which we evaluate explicitly to obtain 
\begin{equation}
Y_{10}^{jj}\left(\hat{n}_{1},\hat{n}_{2}\right)=\frac{i\left(-1\right)^{j}}{4\pi\Omega_{j}^{0}}\sqrt{\frac{3\left(2j+1\right)}{2}}\partial_{\phi_{2}}P_{j}\left(\hat{n}_{1}\cdot\hat{n}_{2}\right),\label{eq:YBSH10}
\end{equation}
where $P_{j}$ represents the Legendre polynomial of degree $j$ (see Appendix \ref{sec:Appendix_Clebsch-Gordan-coefficients}). Substituting Equations (\ref{eq:YBSH10}) and $\mathcal{C}_{1jj\omega;00}^{1}$ from Equation (\ref{eq:C_components_defn}) into Equation (\ref{eq:Kphi_l0}), we obtain 
\begin{align}
K_{\phi,10}\left(r;\mathbf{x}_{1},\mathbf{x}_{2}\right) & =8\sqrt{\frac{3}{4\pi}}\frac{\rho}{r}\sum_{j}\frac{\left(2j+1\right)}{4\pi}\int_{0}^{\infty}\frac{d\omega}{2\pi}\,\omega^{3}P\left(\omega\right)\Im\left[h^{*}\left(\mathbf{x}_{1},\mathbf{x}_{2},\omega\right)\right]\times\nonumber \\
 & \Re\left[\gfnomega{0*}0j\left(r_{\mathrm{obs}},r_{\mathrm{src}}\right)\mathcal{G}_{1jj\omega;00}^{1}\left(r,r_{\mathrm{obs}},r_{\mathrm{src}}\right)\right]\partial_{\phi_{2}}P_{j}\left(\hat{n}_{1}\cdot\hat{n}_{2}\right),\label{eq:Kphi_10}
\end{align}
where the expression for $\mathcal{G}_{1jj\omega;00}^{1}$ in terms of the Green function components is listed in Table \ref{tab:Expressions-for-J}.

\begin{figure}
\includegraphics[scale=0.6]{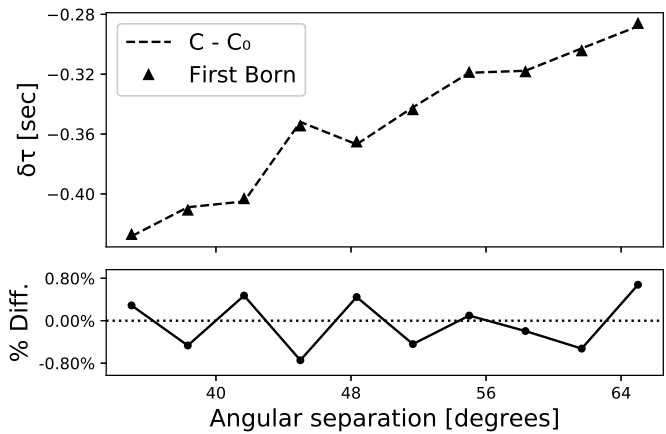}
\caption{Top: Travel-time shifts as a function of  angular separation between two observation points on the equator for waves travelling through a uniformly rotating Sun. The travel-time shifts have been computed: (1) from the difference in the measured cross-covariances (Equation \eqref{eq:dtau_cc}), and (2) by using the first Born approximation (Equation \eqref{eq:dtau_uniform_kernel}). Bottom: Relative difference between the travel times in the top panel.}
\label{fig:Travel-time-shifts}
\end{figure}

We compute the travel times for several observation distances using Equations (\ref{eq:dtau_cc}) and (\ref{eq:dtau_uniform_kernel}), and plot them in Figure \ref{fig:Travel-time-shifts}. The close match between these values serves to validate the sensitivity kernels computed in this work.

\subsection{Numerical evaluation\label{sec:Numerical}}

 We follow a two-step strategy in evaluating the kernel --- at the first step we evaluate the Green function components following \citet{2020ApJ...895..117B} and save them to disk, following which we read the functions in as necessary and compute the kernel using Equation (\ref{eq:K_sepvar}).
The computationally expensive step in the evaluation of the kernel is reading in the pre-computed Green-function FITS files from the
disk, therefore efficient computation of the kernel requires minimizing the number of FITS IO operations. The expression for the kernel in
Equation (\ref{eq:K_sepvar}), while succinct, is not the most convenient form for efficient numerical evaluation. We use Equations (\ref{eq:C_components_defn}) and (\ref{eq:K_components_defn}) to rewrite the expression for the kernel as 
\begin{align}
K_{\gamma,\ell m}\left(r;\mathbf{x}_{1},\mathbf{x}_{2}\right) & =\sum_{j_{1}j_{2}}\sum_{\alpha_{1}\alpha_{2}}\int_{0}^{\infty}\frac{d\omega}{2\pi}\,\omega^{2}P\left(\omega\right)\times\nonumber \\
 & \left(2\Re\left[h^{*}\left(\mathbf{x}_{1},\mathbf{x}_{2},\omega\right)\gfnomega{\alpha_{2}}0{j_{2}}\left(r_{2},r_{\mathrm{src}}\right)J_{\ell j_{1}j_{2}\omega;\alpha_{1}0}^{-\gamma*}\left(r,r_{1},r_{\mathrm{src}}\right)\right]P_{\ell m}^{j_{1}j_{2},\alpha_{1}\alpha_{2}}\left(\mathbf{x}_{1},\mathbf{x}_{2}\right)\right.\nonumber \\
 & \left.+2\Re\left[h^{*}\left(\mathbf{x}_{1},\mathbf{x}_{2},\omega\right)\gfnomega{\alpha_{2}*}0{j_{2}}\left(r_{1},r_{\mathrm{src}}\right)J_{\ell j_{1}j_{2}\omega;\alpha_{1}0}^{-\gamma}\left(r,r_{2},r_{\mathrm{src}}\right)\right]P_{\ell m}^{j_{1}j_{2},\alpha_{1}\alpha_{2}}\left(\mathbf{x}_{2},\mathbf{x}_{1}\right)\right).\label{eq:K_numerical}
\end{align}
Written this way, the Green function component with the source at $r_{\mathrm{src}}$ needs to be read in only for the mode $j_{2}$, whereas the components with the sources at $r_{1}$ and $r_{2}$ needs to be read in for the mode $j_{1}$. Equation (\ref{eq:K_numerical}) appears to come at the expense of an additional computation of the bipolar spherical harmonic $P_{\ell m}^{j_{1}j_{2},\alpha_{1}\alpha_{2}}\left(\mathbf{x}_{2},\mathbf{x}_{1}\right)$, however this might be mitigated to some extent by noting that 
\begin{equation}
    P_{\ell m}^{j_{1}j_{2},\alpha_{1}\alpha_{2}}\left(\mathbf{x}_{2},\mathbf{x}_{1}\right)=\left(-1\right)^{j_{1}+j_{2}+\ell}P_{\ell m}^{j_{2}j_{1},\alpha_{2}\alpha_{1}}\left(\mathbf{x}_{1},\mathbf{x}_{2}\right),
\end{equation}
so we may store the values of $P_{\ell m}^{j_{1}j_{2},\alpha_{1}\alpha_{2}}\left(\mathbf{x}_{1},\mathbf{x}_{2}\right)$ as they are computed, and use pre-computed values of --- if available --- to evaluate $P_{\ell m}^{j_{1}j_{2},\alpha_{1}\alpha_{2}}\left(\mathbf{x}_{2},\mathbf{x}_{1}\right)$
without an explicit loop over the component harmonics.

\begin{figure}
    \includegraphics[scale=0.8]{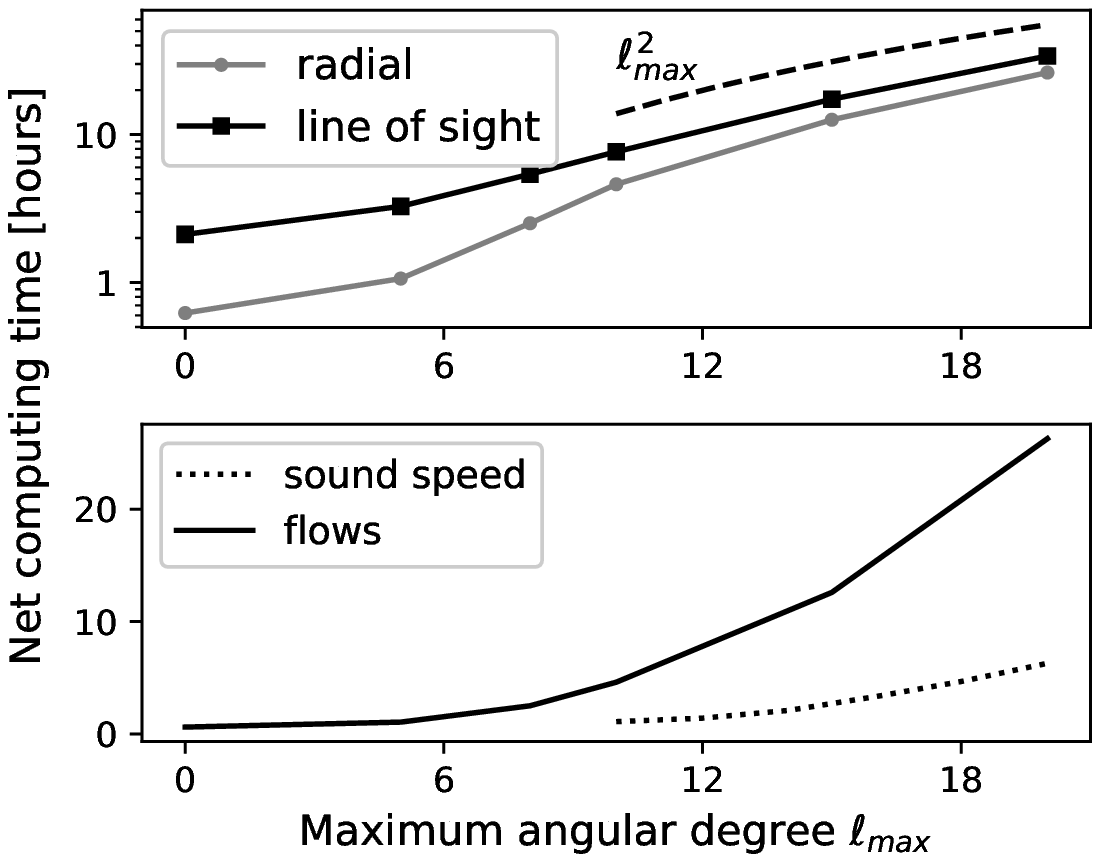}
    \caption{Top: Computational time required to evaluate the kernel components for all modes labelled by $(\ell,m)$ where $|m| \leq \ell$ and $\ell \leq \ell_\mathrm{max}$. The dashed line indicates a scaling $\propto \ell_\mathrm{max}^2$. Bottom: Comparison between the computational time required to compute the kernels for flows (solid line) with that required to compute the kernels for sound speed \citep[][dotted line]{2020ApJ...895..117B}}
    \label{fig:runtimes}
\end{figure}

The computational expense of evaluating the kernel components for all modes is substantial, so this technique is perhaps better suited for large-scale flows where we may restrict the computation to a narrow range of angular degrees. The evaluation time depends on the grid of wave modes used --- both in angular degrees and in temporal frequencies --- as well as the spherical harmonic modes of the flow for which kernels are evaluated. In the present analysis, the code has been written in the Julia programming language \citep{julia}, and is in the form of a map-reduce operation, where the map component --- sums over sections of a range of wave modes and frequencies --- is embarrassingly parallel. We describe the algorithm schematically in Algorithm \ref{algo:kernel}. Given a frequency grid of $N_\nu$ points, a set of $j_\mathrm{max}$ wave modes, a maximum angular degree of $\ell_\mathrm{max}$ and a maximum azimuthal order of $m_\mathrm{max}$ in the PB basis decomposition of the flow velocity, the number of terms that contribute towards the kernel is of the order $\mathcal{O}\left(N_\nu\,j^2_\mathrm{max} \ell_\mathrm{max} m_\mathrm{max}\right)$, where each term involves a sum over radial arrays. Computing the line-of-sight projected kernel would involve summing up four sets of arrays corresponding to $0\leq\alpha_1,\alpha_2\leq 1$, where we use the symmetry relations in Equation \eqref{eq:J_symmetry} to represent the terms corresponding to $\alpha_i=-1$ in terms of $\alpha_i=1$. The time required to read in FITS files from disk may also be reduced by caching the necessary Green-function arrays in memory. We use a grid of frequencies that spans $2.5$ mHz to $4.5$ mHz uniformly over $4000$ points. We also restrict ourselves to wave modes in the range $5 \leq j \leq 80$, where the lower limit arises from the fact that our radial grid does not extend all the way to the center of the Sun, and the upper limit arises from numerical accuracy of the publicly available library SHTOOLS \citep{doi:10.1029/2018GC007529} that we use to compute Clebsch-Gordan coefficients. We carry out the computation on $56$ cores on the Dalma cluster at New York University Abu Dhabi using  2.40GHz Intel Broadwell CPUs. We evaluate the kernel components for all modes $(\ell,m)$ satisfying $\ell \leq\ell_\mathrm{max}$, and we plot the computation time in Figure \ref{fig:runtimes} as a function of $\ell_\mathrm{max}$. The evaluation time required is dominated by FITS input-output operations for low cutoff values of $\ell$, whereas it starts being dominated by the kernel computations for a higher cutoff in $\ell$. This shows up in the reduction in the contrast in evaluation times as the cutoff in $\ell$ increases.

A further optimization might be carried out by noting that the Green functions have power concentrated along distinct ridges corresponding to standing modes in the Sun, therefore suitable filters might eliminate regions of the spectrum that do not contribute significantly to the overall result.

We note that the computational expense involved in this analysis significantly exceeds that required in computing kernels for sound-speed \citep[][]{2020ApJ...895..117B}, as the radial functions involved in evaluating the sound-speed kernel do not depend on $\ell$, whereas for flows, the functions $J_{\ell j_{1}j_{2}\omega;\alpha_{1}0}^{\gamma}\left(r,r_{1},r_{\mathrm{src}}\right)$ need to be re-computed for each $\ell$. We demonstrate the difference in computation time in the bottom panel of Figure \ref{fig:runtimes}, where the time required to compute the kernels for sound-speed are obtained from \citet{2020ApJ...895..117B}.

\begin{figure}
\begin{algorithm}[H]
  \caption{Pseudocode to numerically evaluate the kernel components}
  \label{algo:kernel}
  \algblock[Name]{Parallel}{EndParallel}
   \begin{algorithmic}[1]
   \Function{$K_{\gamma,\ell m}$}{$\mathbf{x}_{1},\mathbf{x}_{2}$,$\omega_\mathrm{array}$,$j_\mathrm{array}$,$\ell_\mathrm{max}$}
   \State Evaluate $h\left(\mathbf{x}_{1},\mathbf{x}_{2},\omega\right)$ and send to all processors
   \State Split $\omega_\mathrm{array}$ and $j_\mathrm{array}$ over the available processors
   \Parallel
        \State $\omega_\mathrm{processor}$ = local section of $\omega_\mathrm{array}$
        \State $j_\mathrm{processor}$ = local section of $j_\mathrm{array}$
        \State Evaluate $P_{\ell m}^{j_{1}j_{2},\alpha_{1}\alpha_{2}}\left(\mathbf{x}_{1},\mathbf{x}_{2}\right)$ and $P_{\ell m}^{j_{1}j_{2},\alpha_{1}\alpha_{2}}\left(\mathbf{x}_{2},\mathbf{x}_{1}\right)$ for necessary parameter values
        
        \State $K_{\gamma,\ell m}$ = 0
    
        \For{$\omega$ in $\omega_\mathrm{processor}$ and $j_2$ in $j_\mathrm{processor}$}
        
            \State Read in $G_{j_2,\omega}(r,r_\mathrm{src})$ and $\partial_r G_{j_2,\omega}(r,r_\mathrm{src})$
        
            \For{$j_1$ in $j_\mathrm{array}$}
                \State Read in $G_{j_1,\omega}(r,r_\mathrm{1})$ and $G_{j_1,\omega}(r,r_\mathrm{2})$
                
                \For{$\alpha_1$ in $0:1$, $\gamma$ in $0:1$ and $i$ in 1:2}
                    \State Evaluate and store individual terms in $\mathcal{G}_{\ell j_{0}j_{2};\alpha_{1}0}^{\gamma}\left(r,r_{i},r_{\mathrm{src}}\right)$ that are independent of $\ell$
                \EndFor
                
                \For{$\ell$ in $0:\ell_\mathrm{max}$}
                
                    \If{$|j_1 - j_2|\leq \ell \leq j_1 + j_2$}
                        \For{$\gamma$ in $0:1$, $\alpha_1$ in $0:1$ and $i$ in 1:2}
                            \State Evaluate and store $J_{\ell j_{1}j_{2}\omega;\alpha_{1}0}^{\gamma}\left(r,r_{i},r_{\mathrm{src}}\right)$ 
                        \EndFor
                        \For{$m$ in $0:\ell$}
                            \State $T_1$ = 0
                            \For{$\gamma$ in $0:1$}
                                \State $T_{\gamma}$ = Sum over $\alpha_1$ and  $\alpha_2$ in Equation \eqref{eq:K_numerical}
                                \State $K_{\gamma,\ell m}$ += $T_{\gamma}$
                            \EndFor
                            \State $K_{-1,\ell m}$ += $(-1)^{\ell+j_1+j_2} T_{1}$
                        \EndFor
                    
                    \EndIf
                \EndFor
            \EndFor
        \EndFor
   \EndParallel
   
   \State $K_{\gamma,\ell m}$ = Sum over $K_{\gamma,\ell m}$ across all processors
   
   \State Return $K_{\gamma,\ell m}$
   
   \EndFunction
   \end{algorithmic}
\end{algorithm}
\end{figure}

\subsection{Exploiting spherical symmetry}

One advantages of a spherical-harmonic decomposition of the kernel is that the transformation of bipolar spherical harmonics on rotation of coordinate systems is well known --- they get coupled to other components with the same degree $\ell$ through the Wigner D-matrix. If a rotation characterized by the Euler angles $(\alpha,\beta,\gamma)$ is carried out to the coordinate frame, the components $P_{\ell,m}(\hat{n}_1,\hat{n}_1)$ of a two-point field on the surface of a sphere in the new coordinate frame are related to those in the old one through

\begin{equation}
P_{\ell m}\left(\hat{n}_{1}^{\prime},\hat{n}_{2}^{\prime}\right)=\sum_{m^{\prime}}D_{m^{\prime}m}^{\ell}\left(\alpha,\beta,\gamma\right)P_{\ell m^{\prime}}\left(\hat{n}_{1},\hat{n}_{2}\right).\label{eq:BSH_rot}
\end{equation}

where the Wigner D-matrix $D^\ell_{m^\prime m}(\alpha,\beta,\gamma)$ acts as the rotation matrix. The relation is valid for tensor spherical harmonics as well, where the non-$m$ indices are carried through unchanged.

\begin{figure}
\includegraphics[scale=0.8]{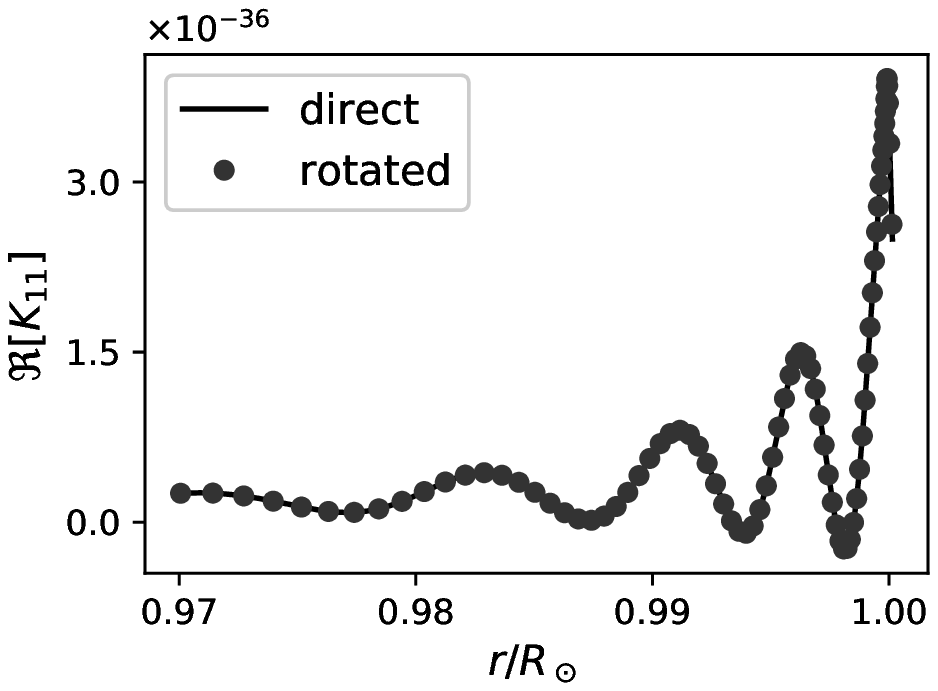}
\caption{Radial profiles of kernel components $K_{11}(r,\mathbf{x}_1,\mathbf{x}_2)$ for $\mathbf{x}_1=(R_\odot+200\,\mathrm{km},\pi/2,0)$ and $\mathbf{x}_1=(R_\odot+200\,\mathrm{km},\pi/2,\pi/4)$, computed directly by using Equation \eqref{eq:kernel_components}., and by rotating that computed for two points shifted by $\pi/10$ along the Equator. The kernels have been computed using the radial components of the wave velocity.}
\label{fig:kernelrot}
\end{figure}

We use this relation to note that the kernel components $K_{\gamma,\ell m}$ need to be evaluated only once for each angular spacing between the two observation points, and subsequently be evaluated for other points that are spaced identically using Equation \eqref{eq:BSH_rot}. We demonstrate the procedure by choosing two sets of points $\hat{n}_1=(\pi/2,0)\,,\hat{n}_2=(\pi/4,0)$, and $\hat{n}_1^\prime=(\pi/2-\pi/10,0)\,,\hat{n}_2^\prime=(\pi/4-\pi/10,0)$, where the latter pair is related to the former by a rotation about the $y$-axis by $\pi/10$ radians, and all observations are assumed to be carried out at a height of $200$ km above the photosphere. To demonstrate the procedure we compute the kernels without assuming line-of-sight projections, but that may be incorporated into the analysis by simultaneously rotating the harmonics as well as the projection vectors. We compute the kernel components in two approaches: (1) by computing the kernel for $(\hat{n}_1,\hat{n}_2)$ and rotating the components using Equation \eqref{eq:BSH_rot} to obtain the components for $(\hat{n}_1^\prime,\hat{n}_2^\prime)$, and (2) by directly evaluating the kernel components for $(\hat{n}_1^\prime,\hat{n}_2^\prime)$. We refer to the former approach as the "rotated" one, whereas the latter is the "direct" computation. We plot the radial profile of the real part of the kernel component $K_{11}(\hat{n}_1^\prime,\hat{n}_2^\prime)$ computed in the two approaches in Figure \ref{fig:kernelrot}. We demonstrate that the two approaches produce identical results, therefore illustrating the promise of such an approach. Additionally such an approach may allow efficient averaging of kernels over arcs in a point-arc measurement configuration, where the angular distance between the observation points stays fixed.

We note that this particular symmetry is useful only in the scenario where we do not consider center-to-limb variations in observation heights. Including this breaks spherical symmetry irreparably, and a full evaluation of the kernel might be necessary.

\section{Conclusion}

We have presented a scheme that may be used to evaluate sensitivity kernels for large-scale flows in the Sun in spherical geometry, while accounting for line-of-sight projection and line-formation heights that leave systematic imprints in the measurements. Further work needs to be carried out to incorporate filters that are used on seismic data to get these kernels to correspond exactly to measurements. Time-distance analysis also usually relies on travel-time differences rather than the point-to-point travel times themselves, but this is easy to incorporate into this analysis scheme.

The computation of the kernels is carried out assuming that the observation heights are different at different points on the Sun. In this paper we have not explored the ramifications of this on the forward problem of estimating travel-times given profiles of subsurface flows, however it might be interesting to check to what extent this contributes to the systematic travel-time shifts observed by \citet{2013ApJ...774L..29Z} and \citet{2015ApJ...808...59K}. The physical origin of the center-to-limb effect is not clear, with effects such as interactions of seismic waves with granulation \citep{2012ApJ...760L...1B,2012SoPh..275..207S,2015ApJ...808...59K} and foreshortening \citep{2016SoPh..291..731Z} also potentially polluting seismic measurements, although, as the authors demonstrate, the latter does not affect travel-time differences significantly. Eliminating certain trends from first principle might help in studying the ones that remain.

The analysis presented here is computationally more efficient than previous attempts to numerically evaluate the full three-dimensional kernel, nevertheless it remains significantly expensive if a large number of modes are simultaneously sought. This approach is more suited to studies where a small range of modes are necessary, such as for large-scale or axisymmetric flows. Fortunately these constitute several classes of flows on the Sun that are of interest. The approach of \citet{2018A&A...616A.156F} relies on a scalar wave equation, therefore it is expected to be more efficient at computing kernels. It will be interesting to compare the trade-off between computational time and accuracy between the two approaches.

\acknowledgments
This work was supported by NYUAD Institute Grant
G1502 "NYUAD Center for Space Science".
This research was carried out on the High Performance Computing resources at New York University Abu Dhabi. 

\appendix{}

\section{VSH triple integral\label{sec:Appendix_VSH-triple-integral}}

We compute the triple integral 
\[
I_{\ell_{2}m_{2}\ell_{1}m_{1}\ell_{3}m_{3}}^{n_{2}n_{1}n_{3}}\left(f_{\ell_{3}m_{3}}^{n_{3}}\left(r\right)\right)=\int d\hat{n}\,\PB{n_{1}}{\ell_{1}m_{1}}\left(\hat{n}\right)\cdot\left[\PB{n_{2}}{\ell_{2}m_{2}}\left(\hat{n}\right)\cdot\grad\left(f_{\ell_{3}}^{n_{3}}\left(r\right)\PB{n_{3}}{\ell_{3}m_{3}}\left(\hat{n}\right)\right)\right],
\]
where we have chosen the ordering of the superscripts and subscripts
keeping a later result in mind.

We use the expression for $\grad\left(f_{\ell_{3}}^{n_{3}}\left(r\right)\PB{n_{3}}{\ell_{3}m_{3}}\left(\hat{n}\right)\right)$
from Equation (\ref{eq:grad_VSH_PB}) and the relation 
\begin{equation}
\PB n{\ell m}\left(\hat{n}\right)\cdot\mathbf{e}_{\alpha}=\left(-1\right)^{\alpha}Y_{\ell m}^{-\alpha}\left(\hat{n}\right)\delta_{n,-\alpha}
\end{equation}
to obtain
\begin{align*}
\PB{n_{1}}{\ell_{1}m_{1}}\left(\hat{n}\right)\cdot\left(\PB{n_{2}}{\ell_{2}m_{2}}\left(\hat{n}\right)\cdot\grad\left(f_{\ell_{3}}^{n_{3}}\left(r\right)\PB{n_{3}}{\ell_{3}m_{3}}\left(\hat{n}\right)\right)\right) & =\left(-1\right)^{n_{3}}\left[\left(\frac{d}{dr}f_{\ell_{3}}^{n_{3}}\left(r\right)\right)\sgsh 0{\ell_{2}m_{2}}\left(\hat{n}\right)\sgsh{-n_{3}}{\ell_{1}m_{1}}\left(\hat{n}\right)\sgsh{n_{3}}{\ell_{3}m_{3}}\left(\hat{n}\right)\delta_{n_{1},-n_{3}}\right.\\
 & -\frac{1}{r}f_{\ell_{3}}^{n_{3}}\left(r\right)\left[\delta_{n_{2},1}\left(\Omega_{\ell}^{n_{3}}Y_{\ell_{2}m_{2}}^{1}\left(\hat{n}\right)Y_{\ell_{1}m_{1}}^{-n_{3}}\sgsh{-1+n_{3}}{\ell_{3}m_{3}}\left(\hat{n}\right)\left(\hat{n}\right)\delta_{n_{1},-n_{3}}\right.\right.\\
 & \left.+Y_{\ell_{2}m_{2}}^{1}\left(\hat{n}\right)Y_{\ell_{1}m_{1}}^{-\left(n_{3}+1\right)}\left(\hat{n}\right)\sgsh{n_{3}}{\ell_{3}m_{3}}\left(\hat{n}\right)\delta_{n_{1},-\left(n_{3}+1\right)}\delta_{n_{3}}^{-1,0}\right)\\
 & +\delta_{n_{2},-1}\left(\Omega_{\ell}^{-n_{3}}Y_{\ell_{2}m_{2}}^{-1}\left(\hat{n}\right)Y_{\ell_{1}m_{1}}^{-n_{3}}\left(\hat{n}\right)\sgsh{1+n_{3}}{\ell_{3}m_{3}}\left(\hat{n}\right)\delta_{n_{1},-n_{3}}\right.\\
 & \left.\left.\left.+Y_{\ell_{2}m_{2}}^{-1}\left(\hat{n}\right)Y_{\ell_{1}m_{1}}^{-\left(n_{3}-1\right)}\left(\hat{n}\right)\sgsh{n_{3}}{\ell_{3}m_{3}}\left(\hat{n}\right)\delta_{n_{1},-\left(n_{3}-1\right)}\delta_{n_{3}}^{0,1}\right)\right]\right],
\end{align*}
where we have used the shorthand notation $\delta_{a}^{b,c}=\delta_{a,b}+\delta_{a,c}$.
We use the triple integral relation for generalized spherical harmonics:
\begin{equation}
\int_{0}^{\pi}d\Omega\,Y_{\ell_{1}m_{1}}^{n_{1}}\left(\hat{n}\right)Y_{\ell_{2}m_{2}}^{n_{2}}\left(\hat{n}\right)Y_{\ell_{3}m_{3}}^{n_{3}}\left(\hat{n}\right)=\eta_{\ell_{1}\ell_{2}\ell_{3}}\left(\begin{array}{ccc}
\ell_{2} & \ell_{1} & \ell_{3}\\
m_{2} & m_{1} & m_{3}
\end{array}\right)\left(\begin{array}{ccc}
\ell_{2} & \ell_{1} & \ell_{3}\\
n_{2} & n_{1} & n_{3}
\end{array}\right),\label{eq:YSH_triple_integral}
\end{equation}
where 
\[
\eta_{\ell_{1}\ell_{2}\ell_{3}}=\sqrt{\frac{\left(2\ell_{1}+1\right)\left(2\ell_{2}+1\right)\left(2\ell_{3}+1\right)}{4\pi}},
\]
\citep[see][]{DahlenTromp}. The relation in Equation (\ref{eq:YSH_triple_integral})
is valid provided $n_{1}+n_{2}+n_{3}=0$, a condition that is guaranteed
for a scalar quantity. The triple integral evaluates to 
\begin{align}
I_{\ell_{2}m_{2}\ell_{1}m_{1}\ell_{3}m_{3}}^{n_{2}n_{1}n_{3}}\left(f_{\ell_{3}m_{3}}^{n_{3}}\left(r\right)\right) & =\eta_{\ell_{1}\ell_{2}\ell_{3}}\left(\begin{array}{ccc}
\ell_{2} & \ell_{1} & \ell_{3}\\
m_{2} & m_{1} & m_{3}
\end{array}\right)\left(-1\right)^{n_{3}}\times\nonumber \\
 & \left[\left(\frac{d}{dr}f_{\ell_{3}}^{n_{3}}\left(r\right)\right)\left(\begin{array}{ccc}
\ell_{2} & \ell_{1} & \ell_{3}\\
0 & -n_{3} & n_{3}
\end{array}\right)\delta_{n_{1},-n_{3}}\delta_{n_{2},0}\right.\nonumber \\
 & -\delta_{n_{2},1}\frac{1}{r}f_{\ell_{3}}^{n_{3}}\left(r\right)\left\{ \Omega_{\ell_{3}}^{n_{3}}\left(\begin{array}{ccc}
\ell_{2} & \ell_{1} & \ell_{3}\\
1 & -n_{3} & -1+n_{3}
\end{array}\right)\delta_{n_{1},-n_{3}}\right.\nonumber \\
 & \left.+\left(\begin{array}{ccc}
\ell_{2} & \ell_{1} & \ell_{3}\\
1 & -n_{3}-1 & n_{3}
\end{array}\right)\delta_{n_{1},-\left(n_{3}+1\right)}\delta_{n_{3}}^{-1,0}\right\} \nonumber \\
 & -\delta_{n_{2},-1}\frac{1}{r}f_{\ell_{3}}^{n_{3}}\left(r\right)\left\{ \Omega_{\ell_{3}}^{-n_{3}}\left(\begin{array}{ccc}
\ell_{2} & \ell_{1} & \ell_{3}\\
-1 & -n_{3} & n_{3}+1
\end{array}\right)\delta_{n_{1},-n_{3}}\right.\nonumber \\
 & \left.\left.+\left(\begin{array}{ccc}
\ell_{2} & \ell_{1} & \ell_{3}\\
-1 & -n_{3}+1 & n_{3}
\end{array}\right)\delta_{n_{1},-\left(n_{3}-1\right)}\delta_{n_{3}}^{0,1}\right\} \right].
\end{align}
We rewrite the integral in terms of Clebsch-Gordan coefficients using
the relation 
\[
C_{\ell_{1}m_{1}\ell_{3}m_{3}}^{\ell_{2}-m_{2}}=\left(-1\right)^{\ell_{1}-\ell_{3}+m_{2}}\sqrt{2\ell_{2}+1}\left(\begin{array}{ccc}
\ell_{2} & \ell_{1} & \ell_{3}\\
m_{2} & m_{1} & m_{3}
\end{array}\right),
\]
to obtain 
\begin{align*}
I_{\ell_{2}m_{2}\ell_{1}m_{1}\ell_{3}m_{3}}^{n_{2}n_{1}n_{3}}\left(f_{\ell_{3}}^{n_{3}}\left(r\right)\right) & =\left(-1\right)^{m_{2}}C_{\ell_{1}m_{1}\ell_{3}m_{3}}^{\ell_{2}-m_{2}}J_{\ell_{2}\ell_{1}\ell_{3}}^{n_{2}n_{1}n_{3}}\left(f_{\ell_{3}}^{n_{3}}\left(r\right)\right),
\end{align*}
where 
\begin{align*}
J_{\ell_{2}\ell_{1}\ell_{3}}^{n_{2}n_{1}n_{3}}\left(f_{\ell_{3}}^{n_{3}}\left(r\right)\right) & =\eta_{\ell_{2}}^{\ell_{1}\ell_{3}}\left(-1\right)^{n_{3}}\left[\delta_{n_{2},0}\left(\frac{d}{dr}f_{\ell_{3}}^{n_{3}}\left(r\right)\right)C_{\ell_{1}-n_{3}\ell_{3}n_{3}}^{\ell_{2}0}\delta_{n_{1},-n_{3}}\right.\\
 & +\delta_{n_{2},1}\frac{1}{r}f_{\ell_{3}}^{n_{3}}\left(r\right)\left\{ \Omega_{\ell_{3}}^{n_{3}}C_{\ell_{1}-n_{3}\ell_{3}n_{3}-1}^{\ell_{2}-1}\delta_{n_{1},-n_{3}}+C_{\ell_{1}-n_{3}-1\ell_{3}n_{3}}^{\ell_{2}-1}\delta_{n_{1},-\left(n_{3}+1\right)}\delta_{n_{3}}^{-1,0}\right\} \\
 & \left.+\delta_{n_{2},-1}\frac{1}{r}f_{\ell_{3}}^{n_{3}}\left(r\right)\left\{ \Omega_{\ell_{3}}^{-n_{3}}C_{\ell_{1}-n_{3}\ell_{3}n_{3}+1}^{\ell_{2}1}\delta_{n_{1},-n_{3}}+C_{\ell_{1}-n_{3}+1\ell_{3}n_{3}}^{\ell_{2}1}\delta_{n_{1},-\left(n_{3}-1\right)}\delta_{n_{3}}^{0,1}\right\} \right],
\end{align*}
where the pre-factor $\eta_{\ell_{2}}^{\ell_{1}\ell_{3}}$ is given
by 
\[
\eta_{\ell_{2}}^{\ell_{1}\ell_{3}}=\sqrt{\frac{\left(2\ell_{1}+1\right)\left(2\ell_{3}+1\right)}{4\pi\left(2\ell_{2}+1\right)}}.
\]
We also evaluate the sum 
\[
H_{\ell_{2}\ell_{1}\ell_{3}}^{\gamma}\left(g_{\ell_{1}},f_{\ell_{3}}\right)\left(r\right)=\sum_{n_{1}n_{3}}g_{\ell_{1}}^{n_{1}}\left(r\right)J_{\ell_{2}\ell_{1}\ell_{3}}^{\gamma n_{1}n_{3}}\left(f_{\ell_{3}}^{n_{3}}\left(r\right)\right),
\]
for functions $f_{j}^{n}\left(r\right)$ and $g_{j}^{n}$ that satisfies
$f_{j}^{-n}\left(r\right)=f_{j}^{n}\left(r\right)$ and $g_{j}^{-n}\left(r\right)=g_{j}^{n}\left(r\right)$.
We use the Clebsch-Gordan relations 
\begin{align*}
\Omega_{\ell_{3}}^{0}C_{\ell_{3}1\ell_{1}0}^{\ell_{2}1}+\Omega_{\ell_{1}}^{0}C_{\ell_{3}0\ell_{1}1}^{\ell_{2}1} & =\Omega_{\ell_{2}}^{0}C_{\ell_{3}0\ell_{1}0}^{\ell_{2}0},\\
C_{\ell_{3}1\ell_{1}-1}^{\ell_{2}0} & =-C_{\ell_{3}0\ell_{1}0}^{\ell_{2}0}\frac{\left(\left(\Omega_{\ell_{1}}^{0}\right)^{2}+\left(\Omega_{\ell_{3}}^{0}\right)^{2}-\left(\Omega_{\ell_{2}}^{0}\right)^{2}\right)}{2\Omega_{\ell_{3}}^{0}\Omega_{\ell_{1}}^{0}},\\
\Omega_{\ell_{3}}^{2}\Omega_{\ell_{1}}^{0}C_{\ell_{3}2\ell_{1}-1}^{\ell_{2}1}+\Omega_{\ell_{3}}^{0}\Omega_{\ell_{1}}^{0}C_{\ell_{3}0\ell_{1}1}^{\ell_{2}1} & =-\left(\left(\Omega_{\ell_{1}}^{0}\right)^{2}+\left(\Omega_{\ell_{3}}^{0}\right)^{2}-\left(\Omega_{\ell_{2}}^{0}\right)^{2}\right)C_{\ell_{3}1\ell_{1}0}^{\ell_{2}1},
\end{align*}
to expand the sum and obtain 
\begin{align*}
H_{\ell_{2}\ell_{1}\ell_{3}}^{0}\left(g_{\ell_{1}},f_{\ell_{3}}\right)\left(r\right) & =\eta_{\ell_{2}}^{\ell_{1}\ell_{3}}C_{\ell_{1}0\ell_{3}0}^{\ell_{2}0}\left[g_{\ell_{1}}^{0}\left(r\right)\frac{d}{dr}f_{\ell_{3}}^{0}\left(r\right)\right.\\
 & \left.+\left(\frac{g_{\ell_{1}}^{1}\left(r\right)}{\Omega_{\ell_{1}}^{0}}\right)\frac{d}{dr}\left(\frac{f_{\ell_{3}}^{1}\left(r\right)}{\Omega_{\ell_{3}}^{0}}\right)\left(\left(\Omega_{\ell_{1}}^{0}\right)^{2}+\left(\Omega_{\ell_{3}}^{0}\right)^{2}-\left(\Omega_{\ell_{2}}^{0}\right)^{2}\right)\right],\\
H_{\ell_{2}\ell_{1}\ell_{3}}^{1}\left(g_{\ell_{1}},f_{\ell_{3}}\right)\left(r\right) & =\eta_{\ell_{2}}^{\ell_{1}\ell_{3}}C_{\ell_{1}0\ell_{3}-1}^{\ell_{2}-1}\frac{\Omega_{\ell_{3}}^{0}}{r}\left[f_{\ell_{3}}^{0}\left(r\right)g_{\ell_{1}}^{0}\left(r\right)-f_{\ell_{3}}^{0}\left(r\right)\left(\frac{g_{\ell_{1}}^{1}\left(r\right)}{\Omega_{\ell_{1}}^{0}}\right)-\left(\frac{f_{\ell_{3}}^{1}\left(r\right)}{\Omega_{\ell_{3}}^{0}}\right)g_{\ell_{1}}^{0}\left(r\right)\right.\\
 & \left.+\left(\frac{f_{\ell_{3}}^{1}\left(r\right)}{\Omega_{\ell_{3}}^{0}}\right)\left(\frac{g_{\ell_{1}}^{1}\left(r\right)}{\Omega_{\ell_{1}}^{0}}\right)\left(\left(\Omega_{\ell_{1}}^{0}\right)^{2}+\left(\Omega_{\ell_{3}}^{0}\right)^{2}-\left(\Omega_{\ell_{2}}^{0}\right)^{2}\right)\right].
\end{align*}

\section{First Born approximation}

\subsection{Green function\label{subsec:Appendix_Green-function}}

We evaluate 
\[
\delta\mathbf{G}\left(\mathbf{x}_{i},\mathbf{x}_{\mathrm{src}};\omega\right)=-\int d\mathbf{x}\mathbf{G}\left(\mathbf{x}_{i},\mathbf{x};\omega\right)\cdot\left[\delta\mathcal{L}\left(\mathbf{x};\omega\right)\mathbf{G}\left(\mathbf{x},\mathbf{x}_{\mathrm{src}};\omega\right)\right],
\]
for $\delta\mathcal{L}\left(\mathbf{x};\omega\right)=2i\omega\rho\mathbf{u}\left(\mathbf{x}\right)\cdot\grad$
and $\mathbf{u}\left(\mathbf{x}\right)=\sum_{\ell m\gamma}u_{\ell m}^{\gamma}\left(r\right)\PBlm{\gamma}\left(\hat{n}\right).$
Substituting these, we obtain 
\[
\delta\mathbf{G}\left(\mathbf{x}_{i},\mathbf{x}_{\mathrm{src}};\omega\right)=-2i\omega\sum_{\ell m\gamma}\int d\mathbf{x}\,\rho u_{\ell m}^{\gamma}\left(r\right)\mathbf{G}\left(\mathbf{x}_{i},\mathbf{x};\omega\right)\cdot\left[\PBlm{\gamma}\left(\hat{n}\right)\cdot\grad\mathbf{G}\left(\mathbf{x},\mathbf{x}_{\mathrm{src}};\omega\right)\right].
\]

We use the expansion of the Green function in the PB VSH from Equation
(\ref{eq:G_PB_VSH}) and the reciprocity relation $\gfnomega{\beta}{\alpha}j\left(r,r_{i}\right)=\gfnomega{\alpha}{\beta}j\left(r_{i},r\right)$
to obtain 
\begin{align}
\delta\mathbf{G}\left(\mathbf{x}_{i},\mathbf{x}_{\mathrm{src}};\omega\right) & =-2i\omega\sum_{\ell m\gamma}\sum_{j_{1}m_{1}\alpha_{1}\beta_{1}}\sum_{j_{2}m_{2}\alpha_{2}\beta_{2}}\PB{\alpha_{1}*}{j_{1}m_{1}}\left(\hat{n}_{i}\right)\PB{\beta_{2}*}{j_{2}m_{2}}\left(\hat{n}_{\mathrm{src}}\right)\times\nonumber \\
 & \int d\mathbf{x}\,\rho u_{\ell m}^{\gamma}\left(r\right)\gfnomega{\beta_{1}}{\alpha_{1}}{j_{1}}\left(r,r_{i}\right)\times\nonumber \\
 & \left[\PB{\beta_{1}}{j_{1}m_{1}}\left(\hat{n}\right)\cdot\left(\PBlm{\gamma}\left(\hat{n}\right)\cdot\grad\gfnomega{\alpha_{2}}{\beta_{2}}{j_{2}}\left(r,r_{\mathrm{src}}\right)\PB{\alpha_{2}}{j_{2}m_{2}}\left(\hat{n}\right)\right)\right].\label{eq:dG_u_VSH}
\end{align}
We use the triple integral relation from Section \ref{sec:Appendix_VSH-triple-integral}
to obtain 
\begin{equation}
\PB{\beta_{1}}{j_{1}m_{1}}\left(\hat{n}\right)\cdot\left(\PBlm{\gamma}\left(\hat{n}\right)\cdot\grad\gfnomega{\alpha_{2}}{\beta_{2}}{j_{2}}\left(r,r_{\mathrm{src}}\right)\PB{\alpha_{2}}{j_{2}m_{2}}\left(\hat{n}\right)\right)=\left(-1\right)^{m}C_{j_{1}m_{1}j_{2}m_{2}}^{\ell-m}J_{\ell j_{1}j_{2}}^{\gamma\beta_{1}\alpha_{2}}\left(\gfnomega{\alpha_{2}}{\beta_{2}}{j_{2}}\left(r,r_{\mathrm{src}}\right)\right).\label{eq:VSH_triple_int_dG}
\end{equation}
Substituting Equation (\ref{eq:VSH_triple_int_dG}) into Equation
(\ref{eq:dG_u_VSH}) we obtain 
\begin{align*}
\delta\mathbf{G}\left(\mathbf{x}_{i},\mathbf{x}_{\mathrm{src}};\omega\right) & =-2i\omega\sum_{\ell m\gamma}\sum_{j_{1}m_{1}}\sum_{j_{2}m_{2}}\sum_{\alpha_{1}\beta_{2}}\left(-1\right)^{m}C_{j_{1}m_{1}j_{2}m_{2}}^{\ell-m}\PB{\alpha_{1}*}{j_{1}m_{1}}\left(\hat{n}_{i}\right)\PB{\beta_{2}*}{j_{2}m_{2}}\left(\hat{n}_{\mathrm{src}}\right)\times\\
 & \int r^{2}dr\,\rho u_{\ell m}^{\gamma}\left(r\right)\sum_{\beta_{1}\alpha_{2}}\gfnomega{\beta_{1}}{\alpha_{1}}{j_{1}}\left(r,r_{i}\right)J_{\ell j_{1}j_{2}}^{\gamma\beta_{1}\alpha_{2}}\left(\gfnomega{\alpha_{2}}{\beta_{2}}{j_{2}}\left(r,r_{\mathrm{src}}\right)\right).
\end{align*}
We simplify the notation by defining 
\[
J_{\ell j_{1}j_{2}\omega;\alpha_{1}\beta_{2}}^{\gamma}\left(r,r_{i},r_{\mathrm{src}}\right)=-2i\omega\rho\,\sum_{\beta_{1}\alpha_{2}}\gfnomega{\beta_{1}}{\alpha_{1}}{j_{1}}\left(r,r_{i}\right)J_{\ell j_{1}j_{2}}^{\gamma\beta_{1}\alpha_{2}}\left(\gfnomega{\alpha_{2}}{\beta_{2}}{j_{2}}\left(r,r_{\mathrm{src}}\right)\right),
\]
to obtain 
\begin{align*}
\delta\mathbf{G}\left(\mathbf{x}_{i},\mathbf{x}_{\mathrm{src}};\omega\right) & =\sum_{\ell m\gamma}\sum_{j_{1}j_{2}}\sum_{m_{1}m_{2}}\sum_{\alpha_{1}\beta_{2}}\left(-1\right)^{m}C_{j_{1}m_{1}j_{2}m_{2}}^{\ell-m}\PB{\alpha_{1}*}{j_{1}m_{1}}\left(\hat{n}_{i}\right)\PB{\beta_{2}*}{j_{2}m_{2}}\left(\hat{n}_{\mathrm{src}}\right)\times\\
 & \int d\mathbf{x}\,\rho u_{\ell m}^{\gamma}\left(r\right)J_{\ell j_{1}j_{2}\omega;\alpha_{1}\beta_{2}}^{\gamma}\left(r,r_{i},r_{\mathrm{src}}\right).
\end{align*}
We recognize the angular term to be a bipolar vector spherical harmonic,
and may rewrite the expression as 
\begin{align*}
\delta\mathbf{G}\left(\mathbf{x}_{i},\mathbf{x}_{\mathrm{src}};\omega\right) & =\sum_{\ell m\gamma}\sum_{j_{1}j_{2}}\sum_{\alpha_{1}\beta_{2}}\int r^{2}dr\,u_{\ell m}^{\gamma}\left(r\right)J_{\ell j_{1}j_{2}\omega;\alpha_{1}\beta_{2}}^{-\gamma}\left(r,r_{i},r_{\mathrm{src}}\right)\PB{j_{1}j_{2}\alpha_{1}\beta_{2}}{\ell m}\left(\hat{n}_{i},\hat{n}_{\mathrm{src}}\right).
\end{align*}

\subsection{Cross-covariance\label{sec:Appendix_source_angle_int}}

The change in cross-covariance is defined as 
\[
\delta\mathbf{C}\left(\mathbf{x}_{1},\mathbf{x}_{2};\omega\right)=\omega^{2}P\left(\omega\right)\int d\Omega_{\mathrm{src}}\,\left[\delta\mathbf{G}_{r}^{*}\left(\mathbf{x}_{1},\mathbf{x}_{\mathrm{src}};\omega\right)\mathbf{G}_{r}\left(\mathbf{x}_{2},\mathbf{x}_{\mathrm{src}};\omega\right)+\left(1\leftrightarrow2\right)^{\dagger}\right],
\]
where the integral is over the angular distribution of the sources.
The Green function and its change are defined in Equations (\ref{eq:G_PB_VSH})
and (\ref{eq:dG_PB_VSH}) as 
\begin{align*}
\delta\mathbf{G}_{r}\left(\mathbf{x}_{i},\mathbf{x}_{\mathrm{src}};\omega\right) & =\sum_{\ell m\gamma}\sum_{j_{1}m_{1}}\sum_{j_{2}m_{2}}\sum_{\alpha_{1}}\left(-1\right)^{m}C_{j_{1}m_{1}j_{2}m_{2}}^{\ell-m}\PB{\alpha_{1}*}{j_{1}m_{1}}\left(\hat{n}_{i}\right)Y_{j_{2}m_{2}}^{*}\left(\hat{n}_{\mathrm{src}}\right)\times\\
 & \int r^{2}dr\,u_{\ell m}^{\gamma}\left(r\right)\,J_{\ell j_{1}j_{2}\omega;\alpha_{1}0}^{\gamma}\left(r,r_{i},r_{\mathrm{src}}\right),\\
\mathbf{G}_{r}\left(\mathbf{x}_{j},\mathbf{x}_{\mathrm{src}},\omega\right) & =\sum_{\alpha_{3}}\sum_{j_{3}m_{3}}\gfnomega{\alpha_{3}}0{j_{3}}\left(r_{j},r_{\mathrm{src}}\right)\PB{\alpha_{3}}{j_{3}m_{3}}\left(\hat{n}_{j}\right)Y_{j_{3}m_{3}}^{*}\left(\hat{n}_{\mathrm{src}}\right).
\end{align*}
We also use the conjugation relation $\left(u_{\ell m}^{\gamma}\left(r\right)\right)^{*}=\left(-1\right)^{m}u_{\ell-m}^{-\gamma}\left(r\right)$.
This leads to
\begin{align*}
\int d\Omega_{\mathrm{src}}\,\delta\mathbf{G}_{r}^{*}\left(\mathbf{x}_{1},\mathbf{x}_{\mathrm{src}};\omega\right)\mathbf{G}_{r}\left(\mathbf{x}_{2},\mathbf{x}_{\mathrm{src}};\omega\right) & =\sum_{\ell m\gamma}\sum_{j_{1}j_{2}}\sum_{\alpha_{1}\alpha_{2}}\sum_{m_{1}m_{2}}C_{j_{1}m_{1}j_{2}m_{2}}^{\ell m}\PB{\alpha_{1}}{j_{1}m_{1}}\left(\hat{n}_{1}\right)\PB{\alpha_{2}}{j_{2}m_{2}}\left(\hat{n}_{2}\right)\times\\
 & \int r^{2}dr\,u_{\ell m}^{\gamma}\left(r\right)\,\gfnomega{\alpha_{2}}0{j_{2}}\left(r_{2},r_{\mathrm{src}}\right)J_{\ell j_{1}j_{2}\omega;\alpha_{1}0}^{-\gamma*}\left(r,r_{1},r_{\mathrm{src}}\right).
\end{align*}
The angular dependence is given by the the bipolar vector spherical
harmonic, defined as 
\[
\PB{j_{1}j_{2},\alpha_{1}\alpha_{2}}{\ell m}\left(\hat{n}_{1},\hat{n}_{2}\right)=\sum_{m_{1}m_{2}}C_{j_{1}m_{1}j_{2}m_{2}}^{\ell m}\PB{\alpha_{1}}{j_{1}m_{1}}\left(\hat{n}_{1}\right)\PB{\alpha_{2}}{j_{2}m_{2}}\left(\hat{n}_{2}\right).
\]
In terms of this, we rewrite the integral as 
\begin{gather*}
\int d\Omega_{\mathrm{src}}\,\delta\mathbf{G}_{r}^{*}\left(\mathbf{x}_{1},\mathbf{x}_{\mathrm{src}};\omega\right)\mathbf{G}_{r}\left(\mathbf{x}_{2},\mathbf{x}_{\mathrm{src}};\omega\right)=\\
\sum_{\ell m\gamma}\sum_{j_{1}j_{2}}\sum_{\alpha_{1}\alpha_{2}}\int r^{2}dr\,u_{\ell m}^{\gamma}\left(r\right)\,\gfnomega{\alpha_{2}}0{j_{2}}\left(r_{2},r_{\mathrm{src}}\right)J_{\ell j_{1}j_{2}\omega;\alpha_{1}0}^{-\gamma*}\left(r,r_{1},r_{\mathrm{src}}\right)\PB{j_{1}j_{2},\alpha_{1}\alpha_{2}}{\ell m}\left(\hat{n}_{1},\hat{n}_{2}\right).
\end{gather*}
The second term may be evaluated analogously to 
\begin{gather*}
\int d\Omega_{\mathrm{src}}\,\mathbf{G}_{r}^{*}\left(\mathbf{x}_{1},\mathbf{x}_{\mathrm{src}};\omega\right)\delta\mathbf{G}_{r}\left(\mathbf{x}_{2},\mathbf{x}_{\mathrm{src}};\omega\right)=\\
\sum_{\ell m\gamma}\sum_{j_{1}j_{2}}\sum_{\alpha_{1}\alpha_{2}}\int r^{2}dr\,u_{\ell m}^{\gamma}\left(r\right)\,\gfnomega{\alpha_{1}*}0{j_{1}}\left(r_{1},r_{\mathrm{src}}\right)J_{\ell j_{2}j_{1}\omega;\alpha_{2}0}^{\gamma}\left(r,r_{2},r_{\mathrm{src}}\right)\PB{j_{1}j_{2},\alpha_{1}\alpha_{2}}{\ell m}\left(\hat{n}_{1},\hat{n}_{2}\right).
\end{gather*}

The change in cross-covariance therefore is
\begin{align*}
\delta\mathbf{C}\left(\mathbf{x}_{1},\mathbf{x}_{2};\omega\right) & =\sum_{\ell m\gamma}\sum_{j_{1}j_{2}}\sum_{\alpha_{1}\alpha_{2}}\int r^{2}dr\,u_{\ell m}^{\gamma}\left(r\right)\,\PB{j_{1}j_{2},\alpha_{1}\alpha_{2}}{\ell m}\left(\hat{n}_{1},\hat{n}_{2}\right)\times\\
 & \left[\gfnomega{\alpha_{2}}0{j_{2}}\left(r_{2},r_{\mathrm{src}}\right)J_{\ell j_{1}j_{2}\omega;\alpha_{1}0}^{-\gamma*}\left(r,r_{1},r_{\mathrm{src}}\right)+\gfnomega{\alpha_{1}*}0{j_{1}}\left(r_{1},r_{\mathrm{src}}\right)J_{\ell j_{2}j_{1}\omega;\alpha_{2}0}^{\gamma}\left(r,r_{2},r_{\mathrm{src}}\right)\right].
\end{align*}
We define 
\[
\mathcal{C}_{\ell j_{1}j_{2}\omega;\alpha_{1}\alpha_{2}}^{\gamma}\left(r,r_{1},r_{2},r_{\mathrm{src}}\right)=\gfnomega{\alpha_{2}}0{j_{2}}\left(r_{2},r_{\mathrm{src}}\right)J_{\ell j_{1}j_{2}\omega;\alpha_{1}0}^{-\gamma*}\left(r,r_{1},r_{\mathrm{src}}\right)+\gfnomega{\alpha_{1}*}0{j_{1}}\left(r_{1},r_{\mathrm{src}}\right)J_{\ell j_{2}j_{1}\omega;-\alpha_{2}0}^{\gamma}\left(r,r_{2},r_{\mathrm{src}}\right)
\]
and obtain 
\[
\delta\mathbf{C}\left(\mathbf{x}_{1},\mathbf{x}_{2};\omega\right)=\sum_{\ell m\gamma}\sum_{j_{1}j_{2}}\sum_{\alpha_{1}\alpha_{2}}\int r^{2}dr\,u_{\ell m}^{\gamma}\left(r\right)\,\mathcal{C}_{\ell j_{1}j_{2}\omega;\alpha_{1}\alpha_{2}}^{\gamma}\left(r,r_{1},r_{2},r_{\mathrm{src}}\right)\PB{j_{1}j_{2},\alpha_{1}\alpha_{2}}{\ell m}\left(\hat{n}_{1},\hat{n}_{2}\right).
\]

\section{Bipolar spherical harmonic for uniform rotation\label{sec:Appendix_Clebsch-Gordan-coefficients}}

We use the Clebsch-Gordan recursion relation 
\begin{equation}
C_{\ell m\ell-m}^{10}=\sqrt{\frac{3}{\ell\left(\ell+1\right)}}mC_{\ell m\ell-m}^{00}.\label{eq:CG10_recursion}
\end{equation}
The bipolar spherical harmonic $\sgsh{\ell\ell}{10}\left(\hat{n}_{1},\hat{n}_{2}\right)$
is defined as 
\begin{equation}
\sgsh{\ell\ell}{10}\left(\hat{n}_{1},\hat{n}_{2}\right)=\sum_{m}C_{\ell m\ell-m}^{10}Y_{\ell m}\left(\hat{n}_{1}\right)Y_{\ell-m}\left(\hat{n}_{2}\right).\label{eq:Y10_defn}
\end{equation}
Substituting Equation (\ref{eq:CG10_recursion}) into Equation (\ref{eq:Y10_defn}),
we obtain 
\begin{equation}
\sgsh{\ell\ell}{10}\left(\hat{n}_{1},\hat{n}_{2}\right)=i\sqrt{\frac{3}{\ell\left(\ell+1\right)}}\partial_{\phi_{2}}\sgsh{\ell\ell}{00}\left(\hat{n}_{1},\hat{n}_{2}\right).
\end{equation}
We use the result 
\begin{equation}
\sgsh{\ell\ell}{00}\left(\hat{n}_{1},\hat{n}_{2}\right)=\left(-1\right)^{\ell}\frac{\sqrt{2\ell+1}}{4\pi}P_{\ell}\left(\hat{n}_{1}\cdot\hat{n}_{2}\right),
\end{equation}
where $P_\ell(x)$ is the Legendre polynomial of degree $\ell$ and argument $x$, to obtain 
\begin{equation}
\sgsh{\ell\ell}{10}\left(\hat{n}_{1},\hat{n}_{2}\right)=\frac{i\left(-1\right)^{\ell}}{4\pi}\sqrt{\frac{3\left(2\ell+1\right)}{\ell\left(\ell+1\right)}}\partial_{\phi_{2}}P_{\ell}\left(\hat{n}_{1}\cdot\hat{n}_{2}\right).
\end{equation}

\bibliographystyle{aasjournal}
\bibliography{references}

\end{document}